\pdfoutput=1
\documentclass[review]{elsarticle}
\makeatletter
\def\ps@pprintTitle{%
 \let\@oddhead\@empty
 \let\@evenhead\@empty
 \def\@oddfoot{\centerline{\thepage}}%
 \let\@evenfoot\@oddfoot}
\makeatother
\biboptions{numbers,sort&compress}
\usepackage{graphicx}
\usepackage{amsfonts, amsmath, amssymb}
\usepackage{mathabx}
\usepackage{booktabs,siunitx}
\usepackage[svgnames,table]{xcolor}
\usepackage[tableposition=above]{caption}
\usepackage{pifont}
\usepackage{bm}
\usepackage{cancel}
\usepackage{caption}
\usepackage{color}
\usepackage{enumerate}
\usepackage{float}
\usepackage{hyperref}
\usepackage{soul}
\usepackage[english]{babel}
\usepackage[margin=1in]{geometry}
\usepackage{mathtools}
\usepackage{multirow}
\usepackage{setspace}
\usepackage{subfigure}
\usepackage{stmaryrd}
\usepackage{lineno}
\usepackage{tabularx}
\DeclareUnicodeCharacter{00B3}{\textsuperscript{3}}
\usepackage[ruled,linesnumbered]{algorithm2e}
\RestyleAlgo{ruled}
\SetKwComment{Comment}{/* }{ */}

\SetCommentSty{mycommfont}
\usepackage[normalem]{ulem}

\DeclareMathAlphabet{\mathpzc}{OT1}{pzc}{m}{it}

\renewcommand \d[1]{{\rm{d}} #1}

\newcommand \DDD [2]{\frac{{\rm D} #1}{{\rm D} #2}}
\renewcommand{\vec}[1]{\bm{\mathrm{#1}}}
\newcommand{\V}[1]{\bm{\mathrm{#1}}}

\newcommand{\bsigma}{\boldsymbol{\sigma}}

\newcommand{\etal}{\emph{et al. }}

\newcommand{\rhop}{\rho_\mathrm{p}}

\newcommand{\ncells}{n_{\textrm{cells}}}

\newcommand{\R}{\mathbf{R}}

\newcommand{\rhozero}{\rho_0}
\newcommand{\rhoone}{\rho_1}
\newcommand{\rhotwo}{\rho_2}

\newcommand{\sigmaone}{\bm{\sigma}_1}
\newcommand{\sigmatwo}{\bm{\sigma}_2}

\newcommand{\pone}{p_1}
\newcommand{\ptwo}{p_2}

\newcommand{\volume}{V}

\newcommand{\tp}{T_{\rm a}}

\def \bnabla{\V{\nabla}}

\def \x{\vec{x}}

\def \n{\vec{n}}
\def \r{\vec{r}}

\def \v{\vec{v}}
\def \vone{\vec{v}_1}
\def \pone{p_1}
\def \vpone{\vec{p}_1}
\def \vptwo{\vec{p}_2}

\def \rhoone{\rho_1}
\def \vonehat{\widehat{\vec{v}}_1}
\def \ponehat{\widehat{p}_1}

\def \vtwo{\vec{v}_2}

\def \iota{\imath}

\def \U{\vec{U}}

\def \G{\vec{G}}

\def \Drho{\vec{D_{\rho}}}

\def \vrho{\vec{\rho}}

\def \R{\vec{R}}
\def \n{\vec{n}}
\def \v{\vec{v}}

\def \dx{\Delta x}
\def \dy{\Delta y}

\def \rhop {\rho_\mathrm{p}}

\def \Domain{\mathcal{D}}
\def \Domainf{\mathcal{D}_{\rm f}}
\def \Domainb{\mathcal{D}_{\rm b}}

\def \presdisp{\vec{d}_1}
\def \pfunc{p}

\def \3half{\frac{3}{2}}

\def \etal{\emph{et al.}}

\def \vb{\vec{v}_{\rm b}}
\def \fb{\vec{f}_\textrm{b}}
\def \fbone{\vec{f}_\textrm{1b}}
\def \fbtwo{\vec{f}_\textrm{2b}}
\def \fhydro{\V{\mathcal{F}}}
\def \thydro{\V{\mathcal{T}}}
\def \perm{\alpha}
\def \vpenalty{\widetilde{\vec{v}}}

\usepackage{acronym}
\acrodef{ARF}[ARF]{acoustic radiation force}

\newcommand{\upperRomannumeral}[1]{\uppercase\expandafter{\romannumeral#1}}

\newcommand{\Revone}[1]{{\color{black}#1}}

\begin{document}
\let\today\relax
\let\underbrace\LaTeXunderbrace
\let\overbrace\LaTeXoverbrace

\begin{frontmatter}


\title{Fully resolved particle dynamics over long time scales in acoustofluidic flows: Theory and numerical simulations}


\author[UNL]{Khemraj Gautam Kshetri\fnref{eq_contrib}}
\author[SDSU]{Amneet Pal Singh Bhalla\corref{mycorrespondingauthor}\fnref{eq_contrib}}
\ead{asbhalla@sdsu.edu}
\author[UNL]{Nitesh Nama\corref{mycorrespondingauthor}}
\ead{nitesh.nama@unl.edu}

\address[UNL]{Department of Mechanical \& Materials Engineering, University of Nebraska--Lincoln, Lincoln, NE}
\address[SDSU]{Department of Mechanical Engineering, San Diego State University, San Diego, CA}
\cortext[mycorrespondingauthor]{Corresponding author}
\fntext[eq_contrib]{These authors contributed equally}

\begin{abstract}

We present a perturbation-based theoretical framework, together with its immersed boundary/volume penalization implementation, for long-time simulations of fully resolved rigid-particle dynamics in acoustically actuated flows. A perturbation formulation separates the fast acoustic time scale from the slow mean particle dynamics, avoiding long-time direct integration of the fully compressible Navier--Stokes equations. The immersed rigid body is represented through a Brinkman penalization term in equations posed over a unified fluid–solid domain.
The coupled fluid–particle system is decomposed into a time-harmonic first-order problem governing the acoustic response and a time-averaged second-order problem governing acoustic streaming and mean particle motion. A central challenge in this decomposition is the treatment of the moving fluid–solid interface: the rigid-body boundary condition must be consistently partitioned between first and second order such that the combined response recovers the motion of the underlying fully coupled compressible system. We derive order-consistent expressions for the structural velocities, comprising rigid-body and Stokes-drift contributions, that enter the penalization terms at both orders and formulate the corresponding first- and second-order rigid-body equations of motion. The second-order problem is driven by forcing terms obtained from the first-order acoustic field and, through force- and torque-free constraints, yields the mean translational and angular velocities required to evolve the particle position and orientation over long times. To reduce errors associated with evaluating forces and torques across the diffuse penalized interface, we employ a contour-integration formulation in which the integration surface is displaced into the surrounding fluid.

For a circular particle in two dimensions, the simulations reveal substantial differences in the magnitude and structure of acoustic streaming around fixed and freely moving particles, highlighting the importance of two-way fluid–structure coupling. The results also show that particle-density effects are more complex than suggested by simplified descriptions based solely on the acoustic contrast factor. The perturbation framework captures the mean particle dynamics using time steps associated with the slow particle-motion scale rather than the much smaller acoustic time scale. Finally, simulations of an elliptical particle demonstrate the ability of the method to capture coupled translational and rotational motion. The proposed approach therefore provides an efficient route to fully resolved, long-time simulations of acoustically driven rigid-particle motion while retaining two-way fluid–structure coupling.

\end{abstract}




\end{frontmatter}

\section{Introduction}

The manipulation of fluids and suspended matter through acoustic waves (commonly referred to as `acoustofluidics') has emerged as a powerful platform for precision particle handling in microfluidic systems~\cite{bose2015role,ding2013surface,laurell2014microscale,rasouli2023acoustofluidics,rufo2022acoustofluidics}. In typical acoustofluidic systems, high-frequency acoustic waves propagate through a fluid-filled microchannel and interact nonlinearly with the viscous fluid and immersed particles. These nonlinear interactions produce not only a high-frequency oscillatory response but also a time-averaged, steady fluid motion known as acoustic streaming, as well as a net force on any suspended particle, known as the \emph{acoustic radiation force}~\cite{kshetri2023basic,bruus2012acoustofluidics7,wiklund2012acoustofluidics,doinikov1994acoustic}. Together, acoustic streaming and the radiation force have been leveraged to drive fluid and particle motion at microscales, enabling a diverse range of applications, including label-free cell handling~\cite{nawaz2015acoustofluidic,antfolk2015acoustofluidic,ding2014cell,li2015standing,guo2015controlling,ahmed2016rotational,ozcelik2016acoustofluidic,xie_probing_2016}, cytometry~\cite{wang2019continuous,wu2021high,nawaz2015acoustofluidic}, microscale propulsion~\cite{dillinger2024steerable,dillinger2021ultrasound,ren20193d,mcneill2020wafer}, continuous-flow mixing~\cite{huang2015acoustofluidic,huang2013acoustofluidic,ozcelik2014acoustofluidic,ahmed2013tunable,bachman2020acoustofluidic,durrer2022robot}, and pumping~\cite{junahuang2014reliable,pavlic2023sharp}.

A substantial body of numerical work has been directed at modeling microscale acoustofluidic phenomena~\cite{nyborg1958acoustic,muller2012numerical,kshetri2025consistent,kshetri2023acoustophoresis,kshetri2025numerical,nama2017acoustic, baasch2019acoustic,pavlic2022influence,pavlic2022interparticle,hahn2015numerical,doinikov1994acoustic,doinikov2017acoustic,nama2014investigation,nama2015numerical,barnkob2018acoustically,chen2018three}. These studies commonly employ a perturbation decomposition of the flow variables that exploits the smallness of the acoustic Mach number to separate the fluid response into a first-order oscillatory component and a second-order, time-averaged component. This results in two sets of linear equations, one governing the oscillatory acoustic field and one governing the steady streaming flow, that can be solved sequentially and are computationally more tractable than a direct numerical simulation of the full nonlinear compressible Navier--Stokes system. A common approach to particle tracking within this framework is to compute the time-averaged acoustic radiation force from the acoustic solution obtained in the absence of the particle, and to subsequently integrate a reduced equation of motion to predict the particle trajectory. While computationally efficient and physically insightful, this approach is inherently one-way coupled: the acoustic and streaming fields are insensitive to the particle's evolving position, so any feedback from particle translation and rotation on the surrounding flow is neglected. Although this approximation has been shown to reproduce experimentally observed trajectories of polystyrene particles, it does not capture the coupled fluid--particle dynamics that can become significant for larger particles, or for particles whose properties deviate substantially from the surrounding fluid, or when transient effects during large-displacement migration are of interest. Alternatively, some prior studies introduce the immersed particle within the simulation framework to account for the fluid--particle coupling, but the particle is often held rigidly pinned at a fixed location, such that the system's response is obtained only at an instant and the particle's dynamics are not considered~\cite{sachs2023behavior,liu2023modeling,liu2023shape}. In another approach, Hahn~\etal~\cite{hahn2015numerical} numerically investigated translational and rotational motion of red blood cells, but the streaming field in that approach was computed in the absence of the particle such that the two-way coupling between particle motion and the acoustic and streaming fields was not resolved.

To address these limitations, researchers have pursued fully resolved, time-dependent direct numerical simulations (DNS), in which the particle geometry and surrounding flow are explicitly resolved and the no-slip and kinematic boundary conditions are enforced while resolving the fast oscillatory motion~\cite{kleischmann2024pairwise,klotsa2007interaction,van2022numerical}. These approaches capture the two-way coupling between particle motion and the surrounding fluid, but at a substantial computational cost. In particular, 
the time step must be sufficiently small to resolve each acoustic oscillation period, even though the particle dynamics and mean flow evolve on much longer time scales. As a result, DNS studies are often restricted to a relatively small number of acoustic cycles, which may be insufficient to capture the long-time evolution of the particle and surrounding flow. 
Importantly, despite the wide adoption of perturbation-based decomposition in acoustofluidic modeling, no existing framework combines such a decomposition with the two-way coupled motion of fully resolved particles. Moreover, although body-fitted grid methods have been used extensively in the acoustofluidics literature~\cite{pavlic2022influence,hahn2015numerical,nama2014investigation,liu2023shape,muller2012numerical}, immersed formulations such as the volume penalization method, which enforce moving solid boundaries on a fixed computational grid without remeshing or explicit imposition of boundary conditions at the fluid–solid interface, remain comparatively underexplored.

In this article, we present a perturbation-based theoretical formulation, together with its volume penalization implementation for simulating the oscillatory and mean dynamics of immersed rigid particles within acoustically actuated microchannels. This work extends our recently proposed volume penalization approach for simulating acoustic streaming in complex microchannels by introducing a moving rigid immersed particle within the microchannel~\cite{gautam2025simulating}. We present a consistent decomposition of both the volume-penalized Navier--Stokes equations and rigid body equations of motion to derive two distinct systems of equations: a first-order system governing the fluid--particle system's oscillatory dynamics, and a second-order, time-averaged system governing the system's mean dynamics. At each time step, the hydrodynamic force and torque experienced by the particle are obtained by performing a contour integration away from the diffused fluid--particle interface. The particle position and orientation are subsequently updated using the particle's mean translational and rotational velocities obtained from the first- and second-order solutions. 

To demonstrate the effectiveness of the proposed fully resolved formulation, we first consider a circular particle undergoing purely translational motion and examine the underlying physics of the coupled fluid–particle system. The simulations reveal substantial qualitative and quantitative differences between the acoustic streaming fields generated around freely moving and pinned particles, underscoring the importance of incorporating particle motion directly into the formulation rather than inferring particle trajectories through post-processing of particle-free solutions. We further show that the influence of particle density on the resulting dynamics is more complex than that captured by the commonly used \emph{acoustic contrast factor}, providing additional motivation for a fully coupled treatment. Despite the disparate time scales associated with the fast acoustic oscillations and slow particle migration, the framework permits particle advancement using time steps orders of magnitude larger than those required to resolve the acoustic field directly, substantially improving computational efficiency. Finally, we demonstrate the applicability of the framework to general geometries by simulating an elliptical particle and capturing its coupled translational and rotational dynamics. The proposed framework thus provides an efficient and broadly applicable approach for simulating immersed rigid-body dynamics in acoustically actuated flows, with potential applications to the numerical investigation of acoustofluidic phenomena such as microscale propulsion and particle assembly.

\section{Governing equations}
\label{sec:governing_equations}
Let $\Domain \subset \mathbb{R}^{N}$ denote a fixed domain in spatial dimensions $N=2$ or $3$, occupied by a compressible fluid containing an immersed solid object.
The interaction of acoustics with a compressible viscous fluid containing immersed objects is governed by the volume-penalized compressible Navier--Stokes equations and the thermodynamic equation of state
\begin{subequations}\label{eq:NS}
\begin{alignat}{2}
&\frac{\partial \rho}{\partial t}+\bnabla \cdot(\rho \v)=0, \label{eq:mass} \\
&\frac{\partial (\rho \v)}{\partial t}+\bnabla \cdot(\rho \v \otimes \v)=\bnabla \cdot \bsigma + \fb, \label{eq:momentum}\\
&p=\pfunc(\rho), \label{eq:state}
\end{alignat}
\end{subequations}
in which $\rho$ denotes the spatially varying mass density, $\v$ is the Eulerian velocity field, $p$ is the pressure, $\bsigma$ is the Cauchy stress tensor, and $\vec{f}_\textrm{b}$ represents the Brinkman penalty force. These equations are defined throughout the domain $\Domain$, which is partitioned into two disjoint subdomains: the fluid region $\Domainf(t)\subset\Domain$ and the immersed-body region $\Domainb(t)\subset\Domain$, satisfying $\Domain=\Domainf(t)\cup\Domainb(t)$.
Assuming a linear viscous (Newtonian) fluid, the Cauchy stress tensor can be written as
\begin{align}
    \bm{\sigma}=-p \V{I}+\mu(\bnabla \v+(\bnabla \v)^\intercal)+\lambda (\bnabla \cdot \v) \V{I},\label{eq:sigma}
\end{align}
in which $\mu$ denotes the spatially varying shear viscosity while $\lambda=\mu_{\rm B}-2\mu/3$ denotes the spatially varying second viscosity, with $\mu_{\rm B}$ being the bulk viscosity. \Revone{Following prior studies~\cite{kshetri2025consistent,nama2014investigation,muller2012numerical}, the second-viscosity contribution is retained in Eq.~\eqref{eq:sigma} to capture viscous attenuation of acoustic waves both within and outside the viscous boundary layer.}
The Brinkman penalization force is expressed as
\begin{equation}
\fb = \frac{\chi}{\perm} (\vpenalty-\v), \label{eq:penalty_force}
\end{equation}
such that it imposes the (Eulerian) velocity $\vpenalty$ of the immersed rigid body onto the fictitious fluid contained within $\Domainb(t)$.
The indicator function $\chi$ is employed to identify the immersed body and satisfies $\chi=1$ in $\Domainb(t)$ and $\chi=0$ elsewhere. The parameter $\perm>0$ controls the strength of the penalization, with smaller values corresponding to a stronger enforcement of the body velocity. In the limit $\perm\rightarrow0$, the kinematic constraint $\v=\vpenalty$ is recovered within the immersed-body region $\Domainb(t)$. Throughout this work, $\perm$ is chosen to be finite but sufficiently small so that the immersed body is represented as an effectively rigid, nearly impermeable region.

The velocity of the immersed rigid body can be expressed as 
\begin{equation}
\label{eq:vrigidbody}
    \vb(\x,t)=\U(t)+\bm\Omega(t) \times (\x-\x_{\rm c}(t)), 
\end{equation}
in which $\x$ is an Eulerian position vector relative to a fixed inertial origin $O$, $\x_{\rm c}$ is the position of the center of mass, $\U=\d \x_{\rm c}/\d t$ is its translational velocity, and $\bm\Omega$ is the angular velocity. 
The structural velocity $\vpenalty$ imposed in Eq.~\eqref{eq:penalty_force} is constructed from $\vb$ at each perturbation order in Sec.~\ref{sec:Eulerian_Lagrangian}. A rigid body immersed in a fluid experiences a net hydrodynamic force $\fhydro$ and torque $\thydro$ due to the surrounding fluid given as
\begin{subequations}\label{eq:FandT}
\begin{alignat}{2}
    \fhydro&=\int_{S(t)} \bsigma\,\n\,\mathrm{d}S,\\
    \thydro&= \int_{S(t)} (\x-\x_{\rm c})\times\bsigma\,\n\,\mathrm{d}S,
\end{alignat}
\end{subequations}
in which $S(t)=\partial \Domainb(t)$ is the fluid--solid interface and $\n$ denotes the unit normal pointing out of the body into the fluid. The translational dynamics of the rigid body are governed by Newton's second law
\begin{equation}
    \label{eq:Newton_law_translation}
    \rhop \volume_{\mathrm{p}} \frac{\d \U }{\d t} = \bm \fhydro,
\end{equation}
in which $\rho_\mathrm{p}$ is the constant density of the solid, $\volume_{\mathrm{p}}$ is its volume, and $\U$ denotes its translational velocity, which is uniform throughout the solid domain.
Similarly, the rotational dynamics of the rigid body are governed by
\begin{equation}
    \label{eq:Newton_law_rotation}
    \frac{\d (\bm{\mathcal{I}}_{\rm c} \bm{\Omega})}{\d t}
    = \thydro,
\end{equation}
in which $\bm{\mathcal{I}}_{\rm c}$ denotes the moment-of-inertia tensor of the rigid body about its center of mass, denoted by the subscript `c'. Combined, Eqs.~\eqref{eq:NS}--\eqref{eq:Newton_law_rotation} govern the dynamics of the fluid and immersed objects under acoustic actuation. The left-hand side of Eq.~\eqref{eq:Newton_law_rotation}
can equivalently be written in the familiar Euler form
\begin{equation}
\label{eq:Euler_rotation}
\frac{\d
    \left(\bm{\mathcal{I}}_{\rm c}\bm{\Omega}\right)}
    {\d t}
    =
    \bm{\mathcal{I}}_{\rm c}\dot{\bm{\Omega}}
    +
    \bm{\Omega}\times
    \left(\bm{\mathcal{I}}_{\rm c}\bm{\Omega}\right).
\end{equation}
Note that Eqs.~\eqref{eq:Newton_law_rotation}
and~\eqref{eq:Euler_rotation} are frame independent: the vector and
tensor quantities, including $\bm{\mathcal I}_{\rm c}$,
$\bm{\Omega}$, and $\thydro$, may be represented either in the
body-fixed frame or in the fixed laboratory frame, provided that all
quantities are represented consistently in the same frame.

\subsection{Perturbation approach}
The governing equations and the associated constraints are expanded via a perturbation expansion, where the primary variables are expressed as an asymptotic series 
\begin{subequations}
\label{eq: perturbation}
\begin{alignat}{2}
& \v=\v_0+\epsilon \v'+\epsilon^2 \v''+\mathcal{O}(\epsilon^3),\\
& p=p_0+\epsilon p'+\epsilon^2 p''+\mathcal{O}(\epsilon^3),\\
& \rho=\rho_0+\epsilon \rho'+\epsilon^2\rho''+\mathcal{O}(\epsilon^3),\\
&\U=\U_0+\epsilon \U'+\epsilon^2 \U''+\mathcal{O}(\epsilon^3),\\
&\bm\Omega = \bm\Omega_0 + \epsilon \bm\Omega'+\epsilon^2 \bm\Omega''+\mathcal{O}(\epsilon^3),
\end{alignat}
\end{subequations}
in which $\epsilon \ll 1$ is a dimensionless perturbation parameter, typically taken as the acoustic Mach number, $\epsilon=\omega d_a/c_0$, with $\omega$, $d_a$, and $c_0$ denoting the angular frequency, acoustic displacement amplitude, and speed of sound, respectively. 
We introduce the notation $A_1=\epsilon A'$ and $A_2=\epsilon^2 A''$ for the primary field $A$, where the subscripts denote the corresponding perturbation order. Here, the zeroth-order fields describe the unperturbed equilibrium state of the system, the first-order fields represent its oscillatory response, and the second-order fields correspond to the mean response. 
Throughout this work, the system is assumed to be initially quiescent, and therefore the zeroth-order velocities are taken to be $\v_0 = \U_0=\bm\Omega_0=\bm{0}$.

Substituting the perturbation expansions into the governing equations (Eqs.~\eqref{eq:NS}--\eqref{eq:penalty_force}) and equating terms of equal order in $\epsilon$, the $\mathcal{O}(\epsilon)$ equations governing the first-order oscillatory fields are obtained as
\begin{subequations} \label{eq:first-order-eqs}
\begin{alignat}{2}
&\frac{\partial \rho_1}{\partial t}+\bnabla \cdot\left(\rho_0 \vone\right)=0,\label{eq:first-mass}\\
&\frac{\partial (\rho_0 \vone)}{\partial t}=\bnabla \cdot \bm{\sigma}_1 + \fbone, \label{eq:first-momentum}\\
&p_1 = c_0^2\rho_1, \label{eq:first-state}\\
&\bm{\sigma}_1=-p_1 \V{I}+\mu(\bnabla \vone+(\bnabla \vone)^\intercal)+\lambda (\bnabla \cdot \vone) \V{I},\label{eq:first-sigma}\\
&\fbone = \frac{\chi}{\perm} (\vpenalty_1 - \vone),\label{eq:first-penalty}
\end{alignat}
\end{subequations}
in which $c_0=\sqrt{\frac{\partial p}{\partial \rho}\big|_{\rhozero}}$ is the speed of sound in the fluid and $\vpenalty_1$ is the first-order structural velocity (see Section~\ref{sec:Eulerian_Lagrangian}). Similarly, expanding Eqs.~\eqref{eq:Newton_law_translation} and \eqref{eq:Newton_law_rotation} using the perturbation expansion, the $\mathcal{O}(\epsilon)$ contribution from the rigid body dynamics is obtained as
\begin{subequations}\label{eq:FO_Newtons_balance}
\begin{alignat}{2}
    &\rho_{\rm p}\volume_\mathrm{p}\frac{\d \U_1}{\d t} = \fhydro_1, \label{eq:FO_Newtons_force_balance}\\
    &\frac{\d(\bm{\mathcal{I}}_{\rm c}\bm\Omega_1)}{\d t} = \thydro_{1} \label{eq:FO_Newtons_torque_balance},
\end{alignat}
\end{subequations}
in which $\fhydro_1$ and $\thydro_{1}$ denote the net first-order hydrodynamic force and torque (evaluated relative to the center of mass) acting on the rigid body, respectively.

Consistent with the harmonic nature of the acoustic excitation, the first-order fields are assumed to be time-periodic with period $\tp$ and angular frequency $\omega=2\pi/\tp$. Accordingly, the first-order solution is expressed as
\begin{equation}
A_1(\x, t) = \hat{A}_1(\x)e^{\iota \omega t} = (A_1^{\rm r} (\x) + \iota A_1^{\rm i} (\x))e^{\iota \omega t},\label{eq:first-order-harmonic}
\end{equation}
in which $A_1$ denotes the first-order fields ($\pone$, $\rhoone$,  $\vone$, $\U_1$, or $\bm\Omega_1$). The superscripts $\rm r$ and $\rm i$ denote the real and imaginary components of the complex-valued spatial amplitude, respectively, and $e^{\iota \omega t}$ denotes the complex exponential function, with the understanding that $\textrm{Re}(A_1(\x, t))$ represents the physically relevant quantities of interest.

Next, we collect the $\mathcal{O}(\epsilon^2)$ terms and subsequently apply the temporal averaging operator,
$
\langle A\rangle=\frac{1}{\tp}\int_{t}^{t+\tp} A\,\d t'$,
over an acoustic period to obtain the governing equations for the second-order mean fields as
\begin{subequations}\label{eq:second-order-eqs}
\begin{alignat}{2}
&\bnabla \cdot \langle\rho_0 \vtwo\rangle=-\bnabla \cdot\langle\rho_1 \vone\rangle,\label{eq:second-mass}\\
&\bnabla \cdot (\langle\bm{\sigma}_2\rangle-\langle\rho_0\vone \otimes \vone\rangle)+ \langle \fbtwo \rangle = 0,\label{eq:second-mom}\\
&\langle\bm{\sigma}_2\rangle=-\langle p_2 \rangle \V{I}+\mu(\bnabla \langle \vtwo \rangle+(\bnabla \langle\vtwo\rangle)^\intercal)+\lambda (\bnabla \cdot \langle\vtwo\rangle) \V{I},\\
&\langle \fbtwo \rangle = \frac{\chi}{\perm} (  \vpenalty_2  - \langle \vtwo \rangle),\label{eq:second-penalty}
\end{alignat}
\end{subequations}
in which $\vpenalty_2$ is the second-order structural velocity (see Section~\ref{sec:Eulerian_Lagrangian}).

 Following the same approach, we gather $\mathcal{O}(\epsilon^2)$ terms in Eqs.~\eqref{eq:Newton_law_translation} and \eqref{eq:Newton_law_rotation} to obtain the mean contribution from the
rigid body equations of motion as
%
\begin{subequations}\label{eq:SO_Newtons_balance}
    \begin{alignat}{2}
        &\left\langle\rho_{\rm p}\volume_{\mathrm{p}}\frac{\d\U_2}{\d t}\right\rangle = \left\langle\fhydro_2\right\rangle,\\
    &\left\langle\frac{\d(\bm{\mathcal{I}}_{\rm c}\bm\Omega_2)}{\d t}\right\rangle = \left\langle\thydro_{2}\right\rangle,
    \end{alignat}
\end{subequations}
in which $\fhydro_2$ and $\thydro_{2}$ are the second-order contributions of the net hydrodynamic force and torque (evaluated relative to the center of mass) on the rigid body, respectively. 

We consider the motion of the immersed solid to be quasi-steady at second order such that Eqs.~\eqref{eq:SO_Newtons_balance} reduce to
\begin{subequations}\label{eq:SO_Newtons_balance_constraint}
    \begin{alignat}{2}
        &\left\langle\bm\fhydro_2\right\rangle=0,\\
    & \left\langle\thydro_{2}\right\rangle=0.
    \end{alignat}
\end{subequations}
%
This modeling choice is similar to Regime I in the analytical study by Xie and Vanneste~\cite{xie2014dynamics}, where the characteristic time scale of acceleration is small in comparison to the time scale of the particle's mean motion~\cite{nama2015numerical,muller2012numerical}.
Further details on the derivation of the first- and second-order equations for the Brinkman-penalized formulation of an acoustically actuated fluid system are provided in our prior work~\cite{gautam2025simulating}.  The primary contribution of the present study lies in the incorporation of the dynamics of a moving immersed rigid body within acoustically actuated fluid systems. 
The introduction of a moving rigid body introduces Eqs.~\eqref{eq:FO_Newtons_balance} at the first order and Eqs.~\eqref{eq:SO_Newtons_balance_constraint} at the second order. 
Eqs.~\eqref{eq:first-order-eqs} and \eqref{eq:FO_Newtons_balance} form a coupled PDE–ODE system for the first-order response, while Eqs.~\eqref{eq:second-order-eqs} and \eqref{eq:SO_Newtons_balance_constraint} form a steady PDE system subject to algebraic force- and torque-free constraints.

\subsection{Boundary conditions}
\label{sec: BC}
The test cases presented in this work concern micro-acoustofluidic devices, where the domain of interest typically comprises a liquid-filled microfluidic channel with or without immersed objects. The system's response to acoustic forcing is modeled by prescribing a known, time-periodic oscillatory motion to one or more boundaries of the domain. The displacement of an oscillating boundary is naturally prescribed in a Lagrangian sense as $\presdisp(\overline{\x},t)$, where $\overline{\x}$ is the position of the oscillatory boundary at rest and $\presdisp$ is $\mathcal{O} (\epsilon)$. However, the fluid is typically modeled via an Eulerian framework such that the velocity is expressed as a function of current position $\x=\overline{\x} + \presdisp(\overline{\x},t)$. 
As such, the no-slip boundary condition at the oscillatory boundary is given as
\begin{align}
    \v (\x,t) = \dot{\presdisp}(\overline{\x},t) \quad \textrm{for} \quad \x=\overline{\x} + \presdisp (\overline{\x},t), \label{eq:noslip}
\end{align}
in which $\dot{\presdisp}$ denotes the material derivative of the boundary displacement. Therefore, prior to the use of the perturbation expansion, the left-hand side of  Eq.~\eqref{eq:noslip} needs to be expanded around the mean boundary position $\overline{\x}$ using Taylor series expansion as
\begin{align}
    \v (\overline{\x},t) + \presdisp (\overline{\x},t) \cdot \bnabla \v (\x,t)\Bigr\rvert_{\x=\overline{\x}}  + \mathcal{O} (\epsilon^3) = \dot{\presdisp}(\overline{\x},t).\label{eq: TS_expansion}
\end{align}
Substitution of the perturbation expansion of the fluid velocity $\v=\v_1+\v_2+\mathcal{O} (\epsilon^3)$ on the left-hand side of the above equation and matching orders of $\epsilon$ on both sides yields boundary conditions for the first- and second-order problems as~\cite{bradley1996acoustic,kshetri2025consistent}
\begin{subequations}
\begin{alignat}{2}
    &\vone (\overline{\x},t) = \dot{\presdisp}(\overline{\x},t) ,\label{eq:FOv_d}\\
    & \langle \vtwo (\overline{\x},t) \rangle=-\v^\textrm{SD}=- \langle \presdisp (\overline{\x},t) \cdot \bnabla \vone (\x,t)\Bigr\rvert_{\x=\overline{\x}} \rangle,\label{eq:SDTS}
\end{alignat}
\end{subequations}
in which $\v^\textrm{SD}$ is the Stokes drift, which links the fluid's Eulerian velocity to its mean Lagrangian velocity~\cite{buhler2014waves,nama2017acoustic}
\begin{equation}
    \langle \v^\textrm{L}\rangle:=\langle \vtwo \rangle + \v^\textrm{SD}.\label{eq:vL}
\end{equation}
Therefore, Eq.~\eqref{eq:SDTS} enforces a zero mean Lagrangian fluid velocity at the oscillating boundary.

\subsection{Kinematics of the oscillating rigid body} \label{sec_kinematics_osc_body}

Let $\overline{\V{x}}_{\rm c}$ denote the
center-of-mass position of the current mean configuration of the rigid body, and $\R$ denote a fixed material coordinate attached to the body. The position of the corresponding material point in the current
mean configuration is written as
\begin{equation}
    \overline{\V{x}}(\R)
    =
    \overline{\x}_{\rm c}
    +
    \overline{\r}(\R),
    \label{eq:mean_position}
\end{equation}
in which $\overline{\r}$ is the radius vector of the material point
relative to the center of mass in the current mean configuration.
Unlike the material coordinate $\R$, the radius vector
$\overline{\r}$ may evolve on the slow time scale as the mean
orientation of the rigid body changes.

The instantaneous position of the same material point is expanded
about the current mean configuration as
\begin{equation}
    \x^{\rm L}(\R,t)
    =
    \overline{\x}(\R)
    +
    \bm{\xi}_1(\R,t)
    +
    \bm{\xi}_2(\R,t)
    +
    \mathcal{O}(\epsilon^3),
    \label{eq:material_position_expansion}
\end{equation}
in which $\bm{\xi}_1$ and $\bm{\xi}_2$ denote the first- and
second-order displacements of the material point relative to its
position in the current mean configuration. Here, the superscript `L' denotes the Lagrangian description of the position.

The center-of-mass position is similarly expanded as
\begin{equation}
    \x_{\rm c}(t)
    =
    \overline{\x}_{\rm c}
    +
    \x_{\rm c1}(t)
    +
    \x_{\rm c2}(t)
    +
    \mathcal{O}(\epsilon^3),
    \label{eq:xc_expansion}
\end{equation}
in which $\x_{\rm c1}$ and $\x_{\rm c2}$ denote the first- and
second-order translational displacements relative to the current mean center of mass. Since the body is rigid, the position of the material point relative
to the instantaneous center of mass may be expanded as
\begin{equation}
    \r^{\rm L}
    =
    \overline{\r}
    +
    \r_1
    +
    \r_2
    +
    \mathcal{O}(\epsilon^3).
    \label{eq:radius_expansion}
\end{equation}
Consequently,
\begin{equation}
    \x^{\rm L}
    =
    \x_{\rm c}+\r^{\rm L},
    \label{eq:exact_position_decomposition}
\end{equation}
and substitution of Eqs.~\eqref{eq:mean_position},~\eqref{eq:xc_expansion}, and~\eqref{eq:radius_expansion} into Eq.~\eqref{eq:exact_position_decomposition}
gives

\begin{equation}
    \x^{\rm L}
    =\overline{\x}
    +
    \left(
        \x_{\rm c1}+\r_1
    \right)
    +
    \left(
        \x_{\rm c2}+\r_2
    \right)
    +
    \mathcal{O}(\epsilon^3).
    \label{eq:position_expanded}
\end{equation}

Comparison with Eq.~\eqref{eq:material_position_expansion} therefore
gives
\begin{subequations}
\begin{align}
    \bm{\xi}_1
    &=
    \x_{\rm c1}+\r_1,
    \label{eq:xi1_definition}
    \\
    \bm{\xi}_2
    &=
    \x_{\rm c2}+\r_2.
    \label{eq:xi2_definition}
\end{align}
\end{subequations}

We next obtain the first- and second-order Lagrangian velocities by
differentiating the position of the material point. Since the mean
configuration is held fixed on the fast acoustic time scale,
\begin{equation}
    \frac{\d \overline{\x}}{\d t}
    = 0.
\end{equation}
Thus,
\begin{align}
    \v^{\rm L}
    &=
    \frac{\d \x^{\rm L}}{\d t} =
    \frac{\d \x_{\rm c1}}{\d t}
    +
    \frac{\d \r_1}{\d t}
    +
    \frac{\d \x_{\rm c2}}{\d t}
    +
    \frac{\d \r_2}{\d t}
    +
    \mathcal{O}(\epsilon^3).
    \label{eq:lagrangian_velocity_from_position}
\end{align}
Using
\begin{equation}
    \frac{\d \x_{\rm c1}}{\d t}
    =
    \U_1,
    \qquad
    \frac{\d \x_{\rm c2}}{\d t}
    =
    \U_2,
    \label{eq:center_velocity_def}
\end{equation}
we obtain
\begin{equation}
    \v^{\rm L}
    =
    \U_1
    +
    \U_2
    +
    \frac{\d \r_1}{\d t}
    +
    \frac{\d \r_2}{\d t}
    +
    \mathcal{O}(\epsilon^3).
    \label{eq:lagrangian_velocity_intermediate}
\end{equation}

The rotational motion of a rigid body satisfies
\begin{equation}
    \frac{\d \r^{\rm L}}{\d t}
    =
    \bm{\Omega}
    \times
    \r^{\rm L}.
    \label{eq:rigid_rotation_kinematics}
\end{equation}
Expanding the angular velocity of the body into first- and second-order contributions
\begin{equation}
    \bm{\Omega}
    =
    \bm{\Omega}_1
    +
    \bm{\Omega}_2
    +
    \mathcal{O}(\epsilon^3),
\end{equation}
and substituting Eq.~\eqref{eq:radius_expansion} into
Eq.~\eqref{eq:rigid_rotation_kinematics} gives
\begin{align}
    \frac{\d \r_1}{\d t}
    +
    \frac{\d \r_2}{\d t}
    ={}&
    \left(
        \bm{\Omega}_1+\bm{\Omega}_2
    \right)
    \times
    \left(
        \overline{\r}
        +
        \r_1+\r_2
    \right)
    +
    \mathcal{O}(\epsilon^3)
    \nonumber\\
    ={}&
    \bm{\Omega}_1\times\overline{\r}
    +
    \bm{\Omega}_2\times\overline{\r}
    +
    \bm{\Omega}_1\times\r_1
    +
    \mathcal{O}(\epsilon^3).
    \label{eq:rotation_order_expansion}
\end{align}
Matching perturbation orders yields
\begin{subequations}
\begin{alignat}{2}
    \frac{\d \r_1}{\d t}
    &=
    \bm{\Omega}_1\times\overline{\r},
    \label{eq:r1_evolution}
    \\
    \frac{\d \r_2}{\d t}
    &=
    \bm{\Omega}_2\times\overline{\r}
    +
    \bm{\Omega}_1\times \r_1.
    \label{eq:r2_evolution}
\end{alignat}
\end{subequations}

Combining Eqs.~\eqref{eq:lagrangian_velocity_intermediate},
\eqref{eq:r1_evolution}, and \eqref{eq:r2_evolution}, the
first-order Lagrangian velocity is
\begin{equation}
    \v_{1}^{\rm L}
    =
    \U_1
    +
    \bm{\Omega}_1
    \times
    \overline{\r}
    \label{eq:v1b_lagrangian}
\end{equation}
and the second-order Lagrangian velocity is
\begin{equation}
    \v_{2}^{\rm L}
    =
    \U_2
    +
    \bm{\Omega}_2
    \times
    \overline{\r}
    +
    \bm{\Omega}_1
    \times
    \r_1.
    \label{eq:v2b_lagrangian}
\end{equation}

The first-order rotational displacement $\r_1$ can be
written in terms of the first-order angular displacement
$\bm{\theta}_1$. The latter is defined through the first-order
rotation kinematics
\begin{equation}
    \frac{\d \bm{\theta}_1}{\d t}
    =
    \bm{\Omega}_1.
    \label{eq:theta_omega_relation}
\end{equation}
Note that, in general, the time derivative of a finite angular
displacement is not simply equal to the angular velocity of the body. However, at first order, the rotation is infinitesimal, such that Eq.~\eqref{eq:theta_omega_relation} holds, as derived in Eq.~\eqref{eq:theta1_Omega1} of~\ref{sec:various_omegas_orientation}.  

Integrating Eq.~\eqref{eq:r1_evolution} gives
\begin{equation}
    \r_1
    =
    \bm{\theta}_1
    \times
    \overline{\r} + \V{c}_1,
    \label{eq:r1_theta}
\end{equation}
in which $\V{c}_1$ is the integration constant.
Consequently,
\begin{equation}
    \v_{2}^{\rm L}
    =
    \U_2
    +
    \bm{\Omega}_2
    \times
    \overline{\r}
    +
    \bm{\Omega}_1
    \times
    \left(
        \bm{\theta}_1
        \times
        \overline{\r} + \V{c}_1
    \right).
    \label{eq:v2b_lagrangian_theta}
\end{equation}

\subsection{Eulerian and Lagrangian velocities of the fluid--solid interface}\label{sec:Eulerian_Lagrangian}
%

Next, we relate the Lagrangian velocity of a material point at the
fluid--solid interface to the Eulerian velocity required by the
volume-penalization formulation. The Eulerian velocity field is denoted by
$\v(\x,t)$, whereas the velocity of the material point at the interface
whose mean position is $\overline{\x}$ is denoted by
$\v^{\rm L}(\overline{\x},t)$. Since the Eulerian coordinate $\x$ denotes a fixed
spatial location, it can be identified with the mean
position $\overline{\x}$ of a material point
when that point occupies $\x$ in the current
mean configuration. By the no-slip condition, these
velocities satisfy the exact kinematic relation
\begin{equation}
    \v\left(\x^{\rm L}(\overline{\x},t),t\right)=\v^{\rm L}(\overline{\x},t).
    \label{eq:eulerian_lagrangian_relation}
\end{equation}
Using Eq.~\eqref{eq:position_expanded} and retaining only first-order displacement (for the purpose of Taylor-series expansion), the Eulerian velocity on the left-hand side of
Eq.~\eqref{eq:eulerian_lagrangian_relation} can be expanded about the
position $\overline{\x}$ of the material point in the current mean
configuration as
\begin{equation}
    \v\left(\overline{\x}+\bm{\xi}_1,t\right)
    =\v(\overline{\x},t)
    +\left(\bm{\xi}_1\cdot\bnabla\right)\v(\overline{\x},t)
    +\mathcal{O}(\epsilon^3).
    \label{eq:vb_taylor}
\end{equation}
Expanding the Eulerian and Lagrangian velocities as
\begin{subequations}
\begin{align}
    \v&=\v_1+\v_2+\mathcal{O}(\epsilon^3),\\
    \v^{\rm L}&=\v^{\rm L}_1+\v^{\rm L}_2+\mathcal{O}(\epsilon^3),
\end{align}
\end{subequations}
and noting that $\bm{\xi}_1=\mathcal{O}(\epsilon)$,
Eq.~\eqref{eq:eulerian_lagrangian_relation} becomes
\begin{equation}
    \v_1(\overline{\x},t)+\v_2(\overline{\x},t)
    +\left(\bm{\xi}_1\cdot\bnabla\right)\v_1(\overline{\x},t)
    =\v^{\rm L}_1+\v^{\rm L}_2+\mathcal{O}(\epsilon^3).
    \label{eq:velocity_matching}
\end{equation}
The first- and second-order Lagrangian velocities of the material point at the interface are given by Eqs.~\eqref{eq:v1b_lagrangian} and~\eqref{eq:v2b_lagrangian_theta}, respectively.   

Therefore, at the first order, Eq.~\eqref{eq:velocity_matching} 
gives the Eulerian velocity to be
prescribed within the current mean body domain as
\begin{equation}
    \vpenalty_1=\U_1+\bm{\Omega}_1\times\overline{\r}.
    \label{eq:v1b_final}
\end{equation}
At the second order, averaging Eq.~\eqref{eq:velocity_matching} over an
acoustic period gives the mean second-order velocity to be prescribed as
\begin{equation}
    \vpenalty_2 
    =\U_2+\bm{\Omega}_2\times\overline{\r}
    +\left\langle\bm{\Omega}_1\times\left(\bm{\theta}_1\times\overline{\r}\right)\right\rangle
    -\left\langle\left(\bm{\xi}_1\cdot\bnabla\right)\v_1\right\rangle.
    \label{eq:v2b_mean_final}
\end{equation}
Note that the contribution of the integration constant $\V{c}_1$ to the penultimate term in Eq.~\eqref{eq:v2b_mean_final} vanishes upon time averaging. The last term in Eq.~\eqref{eq:v2b_mean_final} represents the Stokes drift.  For a body-fitted approach that prescribes the fluid's velocity at the fluid--solid interface via a Dirichlet boundary condition, this term is evaluated at the interface approaching from the fluid side. The fluid particles adjacent to the interface move with the body at all times, such that it is the \emph{mean Lagrangian} velocities of the fluid and of the body that must coincide at the interface at second order. Although the adjacent fluid and solid particles follow the same trajectory and therefore have the same mean Lagrangian velocity, they belong to velocity fields whose normal gradients differ at the interface: within the fluid, the velocity varies across the viscous boundary layer, whereas within the body it varies only through the rigid rotation. Since the Stokes drift depends on the local velocity gradient, the fluid and the body can generally have different Stokes drifts.
The diffuse-interface approach in the current framework imposes a continuous velocity $\vone=\vpenalty_{1}$ within the rigid body (including the diffused interface). Therefore, the velocity gradient (and therefore, Stokes drift) is continuous across the interface and recovers the gradients of the respective velocities within the fluid and the solid region with a smooth transition between the two. Eqs.~\eqref{eq:v1b_final} and~\eqref{eq:v2b_mean_final} specify the structural velocities $\vpenalty_{1}$ and $\vpenalty_{2}$ needed for the penalty force terms in the first- and second-order momentum balances, respectively.

\section{Hydrodynamic torque and force}
\label{sec:force_and_torque}
Here, we derive the expressions for the hydrodynamic torques and forces that govern the particle dynamics at each order through Eqs.~\eqref{eq:FO_Newtons_balance} and~\eqref{eq:SO_Newtons_balance_constraint}. Instead of directly integrating hydrodynamic traction over the surface of the body, a contour integral is performed that avoids evaluating derivatives of the velocity field over the diffused fluid--solid interface.

\subsection{Hydrodynamic torque}
We start from the conservative form of the linear momentum equation
\begin{equation}\label{eq:conservative_momentum}
     \frac{\partial}{\partial t}(\rho \v) + \bnabla\cdot(\rho \v \otimes \v -\bsigma) = 0,
\end{equation}
which is valid pointwise in the fluid domain. The balance of angular momentum about a fixed point $O$ is obtained by taking the cross product of the above equation with the position vector $\x$ measured relative to $O$ as
\begin{equation}
    \x\times \frac{\partial}{\partial t}(\rho \v) + \x\times \bnabla\cdot(\rho \v \otimes \v -\bsigma) = 0.
\end{equation}
Integrating over the instantaneous fluid volume $\volume(t)$, bounded internally by the moving fluid--solid interface $S(t)$ and externally by an arbitrary fixed surface $S_\infty$ enclosing the particle, we obtain
\begin{equation}\label{eq:int_balance}
    \iiint_{\volume(t)} \x\times \frac{\partial}{\partial t}(\rho \v)\,\mathrm{d}V + \iiint_{\volume(t)} \x\times \bnabla\cdot(\rho \v \otimes \v -\bsigma)\,\mathrm{d}V = 0.
\end{equation}
Since $\rho \v \otimes \v-\bsigma$ is symmetric, we use the identity $\x\times(\bnabla\cdot\mathbf{M}) = \bnabla\cdot(\x\times\mathbf{M})$ for a symmetric tensor $\mathbf{M}$, and set $\partial\x/\partial t=\bm{0}$ (since $\x$ is a fixed Eulerian position vector) to obtain
%
\begin{equation}
        \iiint_{\volume(t)} \frac{\partial}{\partial t}(\x\times\rho \v)\,\mathrm{d}V + \iiint_{\volume(t)} \bnabla\cdot\left[\x\times (\rho \v \otimes \v -\bsigma)\right]\mathrm{d}V = 0.
\end{equation}
Applying the Leibniz rule to the first term and the divergence theorem to the second term, we obtain
\begin{equation}\label{eq:leibniz_and_divergence}
\begin{aligned}
    &\frac{\d}{\d t}\iiint_{\volume(t)}(\x\times\rho\v)\,\mathrm{d}V
    - \iint_{S(t)}(\x\times\rho\v)(\v\cdot(-\n))\,\mathrm{d}S
    - \iint_{S_\infty}((\x\times\rho\v)(\v_\infty\cdot\n_\infty))\,\mathrm{d}S 
    + \iint_{S(t)}(\x\times\rho\v)(\v\cdot(-\n))\d S \\ 
    &-\iint_{S(t)}\x\times\bsigma\,(-\n)\,\d S+\iint_{S_\infty}\x\times(\rho \v \otimes \v -\bsigma)\,\n_\infty\,\d S = 0.
\end{aligned}
\end{equation}
in which $\n$ is the unit normal of the particle surface pointing into the fluid, $\v_\infty$ is the velocity of the surface $S_\infty$, and $\n_\infty$ is the outward unit normal of $S_\infty$. In the second term, the velocity of the moving interface $S(t)$ has been replaced by the fluid velocity using the no-slip condition. If the integration surface $S_\infty$ is chosen to be stationary, the third term vanishes. Furthermore, the second and fourth terms cancel identically.
Denoting the hydrodynamic torque about $O$ by $\thydro_O=\iint_{S(t)} \x\times\bsigma\,\n\,\mathrm{d}S$ and rearranging, we obtain
\begin{equation}\label{eq:general_tau}
    \thydro_O = \iint_{S_\infty} \x\times(\bsigma-\rho\v\otimes\v)\n_\infty\,\mathrm{d}S - \frac{\d}{\d t}\iiint_{\volume(t)}\x\times\rho\v\,\mathrm{d}V.
\end{equation}
In the Eulerian framework, it is convenient to express vectors and tensors,
such as $\v$, $\x$, $\n$, and $\bsigma$, in the fixed laboratory frame.
Consequently, the hydrodynamic torque defined in Eq.~\eqref{eq:general_tau} is also naturally expressed in the fixed laboratory frame. 


%

Expanding $\rho,\v,$ and $\bsigma$ in powers of $\epsilon$, Eq.~\eqref{eq:general_tau} gives
\begin{equation}\label{eq:general_tau_expanded}
    \begin{aligned}
        \thydro_{O1} + \thydro_{O2} + \mathcal{O}(\epsilon^3) = &\iint_{S_\infty} \x\times\bigg[\sigmaone+\sigmatwo
        -\left(\rhozero+\rhoone\right)\left(\vone+\vtwo\right)\otimes\left(\vone+\vtwo\right)\bigg]\n_\infty\,\mathrm{d}S  \\
        &-\frac{\d}{\d t}\iiint_{\volume(t)}\x\times(\rhozero+\rhoone+\rhotwo)(\vone+\vtwo)\,\mathrm{d}V+\mathcal{O}(\epsilon^3),
    \end{aligned}
\end{equation}
in which we note that $\x$ denotes a fixed (Eulerian) position in space and therefore does not admit any expansion in powers of $\epsilon$. Collecting terms of $\mathcal{O}(\epsilon)$, we obtain
the first-order torque about $O$ as
\begin{equation}\label{eq:torque_FO}
    \thydro_{O1} = \iint_{S_\infty} \x \times \sigmaone \n_\infty\,\mathrm{d}S - \frac{\d}{\d t}\iiint_{\volume(t)} \x \times\rhozero \vone \,\mathrm{d}V.
\end{equation}
Since the volume integral in Eq.~\eqref{eq:torque_FO} is bounded internally by the fluid--solid interface, the diffuse-interface representation of the interface may introduce inaccuracies in its numerical evaluation. To avoid this issue, we rewrite it as an integral over the entire region $\volume_\infty$ bounded by $S_\infty$, including the immersed solid, and subtract the corresponding integral over the volume occupied by the (moving) solid $
V_s(t)$ to obtain
\begin{equation}\label{eq:torque_FO_split}
    \thydro_{O1} = \iint_{S_\infty} \x \times \sigmaone \n_\infty\,\mathrm{d}S - \Bigg(\frac{\d}{\d t}\iiint_{\volume_\infty} \x \times\rhozero \vone \,\mathrm{d}V - \frac{\d}{\d t}\iiint_{\volume_\mathrm{s}(t)} \x \times\rhozero \vone \,\mathrm{d}V\Bigg).
\end{equation}
Using the result in~\ref{sec_ang_mom} for the time derivative of angular momentum, the last term can be written as
\begin{equation}
\label{eqn_dHdt_x}
\frac{\d}{\d t}\iiint_{\volume_\mathrm{s}(t)} \x \times\rhozero \vone \,\mathrm{d}V = 
\rhozero \volume_\mathrm{p}\, \overline{\x}_{\rm c}\times \frac{\d \V{U}_1}{\d t}
+
\frac{1}{\Tilde{\rho}}\bm{\mathcal{I}}_{\rm c}^{\rm lab} \dot{\bm{\Omega}_1},
\end{equation}
with $\Tilde{\rho} = \rhop/\rho_0$ denoting the ratio of the solid density to the fluid density and where the last term in Eq.~\eqref{eqn_dHdt_O} (see~\ref{sec_ang_mom}) drops out of the first-order torque since it is quadratic in $\bm{\Omega}$
and is therefore $\mathcal{O}(\epsilon^2)$. Substituting Eq.~\eqref{eqn_dHdt_x} and noting that $\x$ and  $\volume_\infty$ are time-independent, Eq.~\eqref{eq:torque_FO_split} can be expressed as
\begin{equation}\label{eq:torque_FO_split2}
    \thydro_{O1} = \iint_{S_\infty} \x\times \sigmaone \n_\infty\,\mathrm{d}S - \iiint_{\volume_\infty} \x \times \rhozero \frac{\partial \vone}{\partial t} \,\mathrm{d}V + \rhozero \volume_\mathrm{p}\, \overline{\x}_{\rm c}\times \frac{\d \V{U}_1}{\d t}+\frac{1}{\Tilde{\rho}}\V{\mathcal{I}}_{\rm c}^{\rm lab} \dot{\bm{\Omega}_1} 
    .
\end{equation}
Within the perturbation framework, in which the rigid-body motion is expanded about its mean configuration, it is natural to evaluate the hydrodynamic torque governing the rigid-body rotation in Eq.~\eqref{eq:Newton_law_rotation} about the mean center-of-mass position. Accordingly, we choose the origin at
$\overline{\x}_{\rm c}$ and replace
$\x-\overline{\x}_{\rm c}$ by $\overline{\r}$, which here denotes the position relative to $\overline{\x}_{\rm c}$ of any point in $\volume_\infty$ or on $S_\infty$. Using the
time-harmonic representation of the first-order quantities
(Eq.~\eqref{eq:first-order-harmonic}) and expressing the time-dependent
laboratory-frame inertia tensor in terms of the time-independent body-frame inertia
tensor using Eq.~\eqref{eqn:Ilab_Ibody},
Eq.~\eqref{eq:torque_FO_split2} reduces to
\begin{equation}
\label{eq:torque_FO_split3}
\thydro_1
=
\iint_{S_\infty}
\overline{\r}\times\sigmaone\n_\infty\,\mathrm{d}S
-
\iota\omega\rhozero
\iiint_{\volume_\infty}
\overline{\r}\times\vone\,\mathrm{d}V
+
\frac{\iota\omega}{\Tilde{\rho}}
\left(
\V{\mathcal R}
\V{\mathcal I}_{\rm c}^{\rm body}
\V{\mathcal R}^{\intercal}
\right)
\bm{\Omega}_1 .
\end{equation}

Eq.~\eqref{eq:torque_FO_split3} allows the first-order hydrodynamic
torque to be computed without evaluating velocity gradients across the diffuse fluid–solid interface, and provides $\thydro_1$ in
Eq.~\eqref{eq:FO_Newtons_torque_balance}. The same reasoning applied to
the linear momentum, using the result in~\ref{sec_lin_mom} with
$\rho=\rhozero$, yields the corresponding term
$\iota\omega\rhozero\U_1\volume_{\rm p}$ in the first-order force in
Sec.~\ref{sec:force}.

For the second-order mean hydrodynamic torque, we collect terms of
$\mathcal{O}(\epsilon^2)$ in Eq.~\eqref{eq:general_tau_expanded} and again choose the origin at the center of mass such that $\x-\overline{\x}_{\rm c} = \overline{\r}$. 
Averaging over an acoustic period, the time derivative of the volume integral in
Eq.~\eqref{eq:general_tau_expanded} has a vanishing average for its
part that is periodic on the fast time scale, while its remaining part,
which varies only on the slow time scale, is neglected, consistent with
the quasi-steady assumption in Eqs.~\eqref{eq:SO_Newtons_balance_constraint}. We
therefore obtain\footnote{We note that the time-averaged torque around the mean center-of-mass position $\overline{\x}_{\rm c}$ and the time-averaged torque around the instantaneous center-of-mass position differ by a term
$\langle\x_{\rm c1}\times\fhydro_1\rangle$ which
vanishes identically, since, by
Eq.~\eqref{eq:FO_Newtons_force_balance} and the time-harmonic form of
the first-order quantities,
$ \x_{\rm c1}\times\fhydro_1 =
    \rhop\volume_\mathrm{p}\,\x_{\rm c1}\times\frac{\d\U_1}{\d t}
    =
    -\omega^2\rhop\volume_\mathrm{p}\,\x_{\rm c1}\times\x_{\rm c1}
    = 0.$
}
\begin{equation}\label{eq:torque_SO}
    \left\langle\thydro_2\right\rangle = \iint_{S_\infty}\overline{\r}\times\left(\left\langle\sigmatwo\right\rangle-\left\langle\rhozero\vone\otimes\vone\right\rangle\right)\n_\infty\,\mathrm{d}S.
\end{equation}


\subsection{Hydrodynamic force}
\label{sec:force}
Following an approach similar to the derivation of the hydrodynamic torque in Eqs.~\eqref{eq:int_balance}--\eqref{eq:general_tau}, the linear momentum equation can be rearranged as
\begin{equation}
        \fhydro = \iint_{S_\infty}\left(\bsigma-\rho\v\otimes\v\right)\n_{\infty}\, \d S 
        -\frac{\d}{\d t}\iiint_{\volume(t)}\rho\v \, \d V. \label{eq:fhydrofull}
\end{equation}
Expanding $\rho,\v,$ and $\bsigma$ in Eq.~\eqref{eq:fhydrofull} in powers of $\epsilon$, and collecting terms at first order, we obtain
\begin{equation}
    \fhydro_1= \iint_{S_\infty}\sigmaone\,\n_\infty\,\d S -\frac{\d}{\d t}\iiint_{\volume(t)} \rhozero \vone \,\mathrm{d}V.
\end{equation}
We follow the same approach as used to transform Eq.~\eqref{eq:torque_FO} to Eq.~\eqref{eq:torque_FO_split3} to avoid potential numerical inaccuracies associated with integration over a volume bounded by a diffused interface. This yields
\begin{equation}\label{eq:force_FO_split_final}
    \fhydro_1= \iint_{S_\infty}\sigmaone\,\n_\infty\,\d S - \iiint_{\volume_\infty}\iota \omega \rhozero \vone  \,\d V + \iota\omega\rho_0\U_1 \volume_\mathrm{p},
\end{equation}
where the last term follows from the  result in~\ref{sec_lin_mom} for the time derivative of linear momentum.
For the second-order time-averaged hydrodynamic force, we collect $\mathcal{O}(\epsilon^2)$ terms in Eq.~\eqref{eq:fhydrofull}. Taking the time average over one acoustic period, the last term in Eq.~\eqref{eq:fhydrofull} has a vanishing average for its part that is periodic on the fast time scale, while its remaining part varies on the slow time scale and is neglected, consistent with the quasi-steady assumption in Eqs.~\eqref{eq:SO_Newtons_balance_constraint}. Therefore, we obtain
\begin{equation}\label{eq:force_SO}
    \langle \fhydro_2 \rangle = \iint_{S_\infty} (\langle \sigmatwo \rangle-\langle \rhozero\vone\otimes\vone\rangle)\n_\infty\,\mathrm{d}S.
\end{equation}
A convergence study of the forces and torque with integration contours $S_\infty$ at different radial distances from the fluid--solid interface is presented in~\ref{app:fo_contour_integration}. The results confirm that both quantities become independent of the choice of $S_\infty$ when it is located outside the smeared fluid-solid interface.

\section{Discretized equations of motion}
\subsection{Spatial discretization}
\label{sec_spatial_descritization}
\begin{figure}
    \centering
    \includegraphics[width=0.75\linewidth]{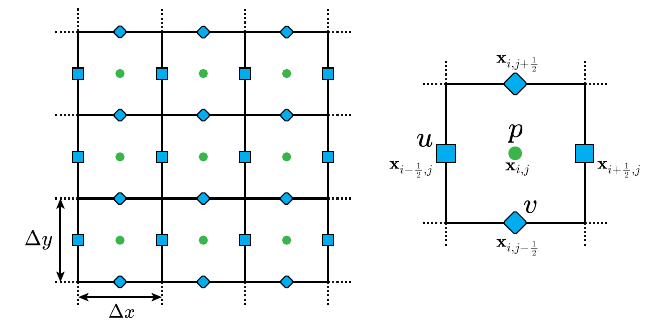}
    \caption{Schematic representation of a 2D staggered Cartesian grid. The left panel shows the staggered-grid coordinate system, with $\Delta x$ and $\Delta y$ denoting the grid spacing in the $x$- and $y$-directions, respectively. The right panel shows an individual grid cell, with the first- and second-order velocity components $u$ ({\color{cyan}$\blacksquare$}) and $v$ ({\color{cyan}$\blackdiamond$}) located at the corresponding side centers, while the scalar pressure is defined at the cell center ({\color{green}$\bullet$}).}
    \label{fig:grid_schematic}
\end{figure}

We discretize the continuous equations of motion on a uniform staggered Cartesian grid by partitioning the computational domain $\Domain$ into $N_x \times N_y$ rectangular cells with spacings $\dx$ and $\dy$ in the $x$- and $y$-directions, respectively, as illustrated in Fig.~\ref{fig:grid_schematic}(left). A uniform grid spacing $\dx = \dy = h$ is used for all numerical results presented in this work.

The $x$-components of the first- and second-order velocities, $u_1$ and $u_2$, are stored at the side centers in the $x$-direction, while the corresponding $y$-components, $v_1$ and $v_2$, are stored at the side centers in the $y$-direction. The first- and second-order pressures, $p_1$ and $p_2$, and all material properties (density $\rhozero$, shear viscosity $\mu$, and second viscosity $\lambda$) are stored at the cell centers. The staggered arrangement is illustrated in Fig.~\ref{fig:grid_schematic}(right). Second-order accurate interpolation and finite-difference schemes are used for the spatial discretization. Details of the discrete operators are provided in our previous work~\cite{gautam2025simulating}.

\subsection{Matrix form of the equations} \label{sec_matrix}
Substituting the time-periodic solution (Eq.~\eqref{eq:first-order-harmonic}) into the first-order mass balance Eq.~\eqref{eq:first-mass} gives
\begin{equation} \label{eq:first-mass-vec} 
 \iota \omega (\rhoone^{\rm r} + \iota \rhoone^{\rm i}) + \bnabla \cdot \left [\rho_0 (\vone^{\rm r} + \iota \vone^{\rm i}) \right] = 0.
\end{equation}
Using $\rhoone^{\rm r}=\pone^{\rm r}/c_0^2$ and $\rhoone^{\rm i}=\pone^{\rm i}/c_0^2$ and separating the real and imaginary parts gives
\begin{subequations} \label{eq:first-mass-comps}
\begin{alignat}{2}
    \textbf{Imaginary:}\quad \bigg(\frac{\omega}{c_0^2}\bigg) \pone^{\rm r} + \bnabla \cdot (\rho_0 \vone^{\rm i}) &= 0, \label{eq:im_cont_first}\\
    \textbf{Real:}\quad -\bigg(\frac{\omega}{c_0^2}\bigg) \pone^{\rm i} + \bnabla \cdot (\rho_0 \vone^{\rm r}) &= 0. \label{eq:re_cont_first}
\end{alignat}
\end{subequations}

Similarly, substituting the time-periodic solution into the first-order momentum Eq.~\eqref{eq:first-momentum} gives
\begin{equation}
\begin{split}
    \iota \omega \rho_0 ( \vone^{\rm r} + \iota \vone^{\rm i} )  = - &\bnabla(\pone^{\rm r} + \iota \pone^{\rm i}) + \bnabla \cdot [\mu (\bnabla ( \vone^{\rm r} + \iota \vone^{\rm i} ) + (\bnabla( \vone^{\rm r} + \iota \vone^{\rm i} ))^\intercal) ]\\
    &+ \bnabla \left [\lambda \bnabla \cdot ( \vone^{\rm r} + \iota \vone^{\rm i} ) \right]+\frac{\chi}{\perm}\left[(\vpenalty_1^{\rm r}+\iota \vpenalty_1^{\rm i}) - (\vone^{\rm r}+\iota \vone^{\rm i})\right].    
\end{split}
\end{equation}
Separating the real and imaginary parts yields
\begin{subequations} \label{eq:first-momentum-comps}
\begin{alignat}{2}
    \textbf{Imaginary:}\quad  \omega \rho_0 \vone^{\rm r} &= -\bnabla \pone^{\rm i}+\bnabla \cdot[\mu(\bnabla \vone^{\rm i}+(\bnabla \vone^{\rm i})^\intercal)]+\nabla[\lambda \bnabla \cdot \vone^{\rm i}]+\frac{\chi}{\perm}(\vpenalty_1^{\rm i}- \vone^{\rm i}), \\
    \textbf{Real:}\quad -\omega \rho_0 \vone^{\rm i} &= -\bnabla \pone^{\rm r}+\bnabla \cdot[\mu(\nabla \vone^{\rm r}+(\bnabla \vone^{\rm r})^\intercal)]+\nabla[\lambda \bnabla \cdot \vone^{\rm r}]+\frac{\chi}{\perm}(\vpenalty_1^{\rm r} - \vone^{\rm r}).
\end{alignat}
\end{subequations}
Eqs.~\eqref{eq:first-momentum-comps} and~\eqref{eq:first-mass-comps} can be written in matrix form as
\begin{equation} \label{eq:matrix-first-brinkman}
\begin{bmatrix}
\omega \vrho_0 &  \V{\chi} / \perm + \V{L_{\mu+\lambda}}  & \V{0} & \G\\
\V{\chi} / \perm + \V{L_{\mu+\lambda}}  & -\omega \vrho_0 & \G & \V{0}\\
\V{0} & \Drho_0 & \frac{\omega}{c_0^2}\V{I} & \V{0}\\
\Drho_0 & \V{0} & \V{0} & -\frac{\omega}{c_0^2}\V{I}
\end{bmatrix}  
\begin{bmatrix}
\vone^{\rm r} \\
\vone^{\rm i} \\
\vpone^{\rm r} \\
\vpone^{\rm i}
\end{bmatrix} 
=
\frac{\V \chi}{\perm} \begin{bmatrix}
\vpenalty_1^{\,\rm i} \\ 
\vpenalty_1^{\,\rm r} \\
\V{0} \\
\V{0}
\end{bmatrix} .
\end{equation}
Here, $\vrho_0$ and $\V{\chi}$ denote the zeroth-order density field and indicator function, respectively, interpolated to the side centers. $\vpenalty_1$ is the first-order structural velocity within the solid (including the diffused interface) region given by Eq.~\eqref{eq:v1b_final}. The discrete spatial operators in Eq.~\eqref{eq:matrix-first-brinkman} are defined as
\begin{subequations} 
\begin{alignat}{2}
    & \Drho_0(\v)=\bnabla \cdot (\rho_0 \v),\\
    & \V{L_{\mu+\lambda}}(\v) 
    =-\bnabla \cdot [\mu (\bnabla \v+(\bnabla \v)^\intercal)+\lambda (\bnabla \cdot \v) \V{I})],\\
    & \V{G}(\V{p})= \bnabla p.
\end{alignat}
\end{subequations}
Eq.~\eqref{eq:matrix-first-brinkman} is solved in conjunction with the first-order constraint Eq.~\eqref{eq:FO_Newtons_balance}, which in matrix form reads as
\begin{align}\label{eq:FO_constraint_real_imag}
    \begin{bmatrix}
        \omega \rhop\volume_\mathrm{p} & 0 & 0 & 0 \\
        0 & -\omega \rhop \volume_\mathrm{p} & 0 & 0 \\
        0 & 0 & \omega \bm{\mathcal{I}}_c^{\rm lab} & 0 \\
        0 & 0 & 0 & -\omega \bm{\mathcal{I}}_c^{\rm lab}
    \end{bmatrix}
    \begin{bmatrix}
        \U_1^{\rm r} \\ \U_1^{\rm i}\\ \bm\Omega_1^{\rm r} \\\bm\Omega_1^{\rm i}
    \end{bmatrix}
    =
    \begin{bmatrix}
        \fhydro_1^{\,\rm i} \\
        \fhydro_1^{\,\rm r} \\
        \thydro_1^{\,\rm i} \\
        \thydro_1^{\,\rm r}
    \end{bmatrix}.
\end{align}
In Eq.~\eqref{eq:FO_constraint_real_imag}, the variation of the
laboratory-frame moment-of-inertia tensor
$\bm{\mathcal{I}}_c^{\rm lab}$ over the fast acoustic time scale does not contribute at first order (since it is multiplied by $\bm\Omega_1^2$; see Eq.~\eqref{eq:Euler_rotation}). At each slow-time configuration,
$\bm{\mathcal{I}}_c^{\rm lab}$ is obtained from the time-independent
body-frame inertia tensor $\bm{\mathcal{I}}_c^{\rm body}$ through the
rotation transformation given in Eq.~\eqref{eqn:Ilab_Ibody}. 


At second order, the governing equations reduce to a steady, low-Mach-number, Brinkman-penalized Stokes system, written as: 
\begin{align} \label{eq:matrix-second-brinkman}
\begin{bmatrix}
\V{\chi} / \perm + \V{L_{\mu+\lambda}} & \G\\
-\Drho_0 & \V{0}
\end{bmatrix}  
\begin{bmatrix}
\vtwo \\
\vptwo
\end{bmatrix} 
&=
\begin{bmatrix}
-\langle \Drho_0(\vone \otimes \vone)\rangle \\
\Drho_0(\v^{\rm SD})
\end{bmatrix}
+
\frac{\V{\chi}}{\perm}
\begin{bmatrix}
\vpenalty_2 \\
\V{0}
\end{bmatrix},
\end{align}
where we have reformulated the source term in the second-order mass balance equation (Eq.~\ref{eq:second-mass}) in terms of $\v^\textrm{SD}$~\cite{kshetri2025consistent}.
In Eq.~\eqref{eq:matrix-second-brinkman}, $\vpenalty_2$ is the prescribed structural velocity within the solid and diffused interface regions as written in Eq.~\eqref{eq:v2b_mean_final}.
Similarly, Eq.~\eqref{eq:matrix-second-brinkman} is solved in conjunction with the second-order constraints, Eqs.~\eqref{eq:SO_Newtons_balance_constraint}.



For the first- and second-order momentum equations, the penalty parameter ${\perm}^{-1}$ scales as $\omega \rhozero$ and $(\mu+\lambda)/h^2$, respectively. In practice, a stronger penalty is used to enforce the velocity boundary condition. Specifically, ${\perm}^{-1}$ is set to $p_\perm \omega \rhozero$ for the first-order system and $p_\perm (\mu+\lambda)/h^2$ for the second-order system, where $p_\perm$ is a case-specific \emph{penalty factor}.

\subsection{Specification of Brinkman penalized zone and contour integration of forces and torques}
\label{sec:penalized_zone}
The solid region $\Domainb$ is represented using a smoothed Heaviside function $\widetilde{H}$, with the indicator function defined as $\chi=1-\widetilde{H}$. The transition occurs over a thickness of $2 \ncells h$ across the fluid--solid interface:
\begin{equation}
    \widetilde{H}_{i, j} = \begin{cases} 0, & \phi_{i, j}<-\ncells h \\
    \frac{1}{2}\left(1+\frac{\phi_{i,j}}{\ncells h}+\frac{1}{\pi}\sin\left(\frac{\pi\phi_{i,j}}{\ncells h}\right)\right), & |\phi_{i,j}|\leq \ncells h \\
    1, & \text{otherwise}.
    \end{cases}
\end{equation}
Here, $\phi$ is a signed-distance function, negative in $\Domainb$ and positive in $\Domainf$, $\ncells$ is the number of smearing cells and $h = \Delta x = \Delta y$ is the unit cell size.

The hydrodynamic force $\fhydro$ and torque $\thydro$  are evaluated on a fixed contour $S_\infty$ enclosing the immersed object as described in Sec.~\ref{sec:force_and_torque}. An intermediate contour $S_\textrm{int}$ is defined by the zero level set of $\psi$ and represented on the Cartesian grid by a stair-step contour $S_\textrm{stair}$, whose faces separate cells with opposite signs of $\psi$. The hydrodynamic force and torque are then computed as
\begin{subequations}
\begin{align}
    \fhydro_{m} &= \sum_{\mathrm{f} \in S_\textrm{stair}} \Bigl[ -p_{m}\n_\mathrm{f}+\mu(\nabla \v_{m}+(\nabla \v_{m})^\intercal)\cdot\n_\mathrm{f}+\lambda (\nabla \cdot \v_{m})\n_\mathrm{f}- \delta_{m2}\langle\rho_0\v_{m-1}(\v_{m-1}\cdot \n_\mathrm{f})\rangle \Bigr]\Delta A_\mathrm{f} \nonumber \\
    &\quad - \delta_{m1}\iota\omega\rho_0 \left[\left( \sum_{\mathrm{c}\,\in \volume_\infty} \v_{m}\,\Delta \volume_\mathrm{c}\right) - \U_{m}\,\volume_\mathrm{p} \right],\\[1ex]
    \thydro_{m} &= \sum_{f \in S_\textrm{stair}} (\x_\mathrm{f}-\overline{\x}_\mathrm{c})\times\Bigl[ -p_{m}\n_\mathrm{f}+\mu(\nabla \v_{m}+(\nabla \v_{m})^\intercal)\cdot\n_\mathrm{f} +\lambda (\nabla \cdot \v_{m})\n_\mathrm{f}- \delta_{m2} \langle\rhozero\v_{m-1}(\v_{m-1}\cdot \n_\mathrm{f})\rangle \Bigr]\Delta A_\mathrm{f}\nonumber \\
    &\quad - \delta_{m1}\iota \omega \rhozero \left( \sum_{\mathrm{c}\,\in \volume_\infty}(\x_\mathrm{f}-\overline{\x}_\mathrm{c})\times \v_{m}\,\Delta \volume_\mathrm{c} \right) + \delta_{m1}\frac{\iota \omega}{\Tilde{\rho}}\left(
    \V{\mathcal R}
    \V{\mathcal I}_{\rm c}^{\rm body}
    \V{\mathcal R}^{\intercal}
    \right)
    \bm{\Omega}_{m}, \label{eq:torque}
\end{align}
\end{subequations}
in which $\n$ is the unit normal, $m = 1,2$ is the perturbation order, $\delta_{mn}$ is the kronecker delta function,  and subscripts $\mathrm{c}$ and $\mathrm{f}$ denote cell and face centers, respectively. $\Delta A_\mathrm{f}$ and $\Delta \volume_\mathrm{c}$ denote area of the cell face and volume of the cell, respectively.
Cell-centered quantities are interpolated to faces by arithmetic averaging. Since the integration surface can be any fixed closed contour enclosing the object, the stair-step representation introduces no additional modeling approximation. Further details of the integration on a Cartesian grid by a stair-step contour are presented in our previous work~\cite{gautam2025simulating}.

\subsection{Rigid body position, angular velocity, and orientation update} \label{sec_RBD_update}

We use a first-order accurate Euler scheme to update the center-of-mass position, angular velocities, and
orientation of the rigid body. The center-of-mass
position is advanced using the second-order translational velocity $\U_2$,
whereas the orientation is advanced using the second-order mean angular
velocity $\V{\Omega}_{\rm mean}$. The latter is expressed in terms of
$\V{\Omega}_1$ and $\V{\Omega}_2$ as
\begin{equation}
    \boldsymbol{\mathrm{\Omega}}_{\mathrm{mean}}
    =
    \boldsymbol{\mathrm{\Omega}}_2
    +
    \frac{1}{2\omega}
    \boldsymbol{\mathrm{\Omega}}_1^{\rm r}
    \times
    \boldsymbol{\mathrm{\Omega}}_1^{\rm i}.
    \label{eq:mean_angular_velocity_main}
\end{equation}
A detailed derivation of Eq.~\eqref{eq:mean_angular_velocity_main} is provided in~\ref{sec:various_omegas_orientation}. In two dimensions, the second
term on the right-hand side of Eq.~\eqref{eq:mean_angular_velocity_main}
vanishes. Note that the position and orientation of the body evolve on the slow time scale. Consequently, the fast oscillatory motion associated with the acoustic time scale is not explicitly resolved when advancing the mean position and orientation, but its effects are incorporated through $\U_2$ and $\boldsymbol{\mathrm{\Omega}}_{\mathrm{mean}}$.

Let $\Delta t$ denote the time-step size and $n$ the current time-step
index. The center-of-mass position is updated according to
\begin{equation}
    \overline{\x}_{\rm c}^{\,n+1}
    =
    \overline{\x}_{\rm c}^{\,n}
    +
    \U_2^{\,n}\Delta t.
    \label{eq:update_xc_eq}
\end{equation}

The mean orientation of the body is represented by the unit quaternion
$\overline{\mathbf{q}}^{\,n}$. Throughout, angular velocities, including
$\V{\Omega}_{\rm mean}$, and hydrodynamic torques are expressed in the
fixed laboratory frame. In contrast, the inertia tensor
$\bm{\mathcal{I}}_{\rm c}^{\rm body}$ is defined in the body-fixed frame and is
therefore time independent.

Assuming that $\V{\Omega}_{\rm mean}$ is constant over a time step, the
incremental rotation angle is 
\begin{equation}
    \Delta\theta^{\,n}
    =
    \Delta t
    \left\|
    \boldsymbol{\mathrm{\Omega}}_{\mathrm{mean}}^{\,n}
    \right\|.
\end{equation}
For
$\|\boldsymbol{\mathrm{\Omega}}_{\mathrm{mean}}^{\,n}\| \ne 0$, the
corresponding incremental quaternion is
\begin{equation}
    \Delta\mathbf{q}^{\,n}
    =
    \left(
    \cos\frac{\Delta\theta^{\,n}}{2},
    \frac{
    \boldsymbol{\mathrm{\Omega}}_{\mathrm{mean}}^{\,n}
    }{
    \|\boldsymbol{\mathrm{\Omega}}_{\mathrm{mean}}^{\,n}\|
    }
    \sin\frac{\Delta\theta^{\,n}}{2}
    \right).
\end{equation}
Because the angular velocity is expressed in the fixed laboratory frame,
the incremental rotation is applied by left multiplication,
\begin{equation}
    \overline{\mathbf{q}}^{\,n+1}
    =
    \Delta\mathbf{q}^{\,n}
    \odot
    \overline{\mathbf{q}}^{\,n},
    \label{eq:mean_quaternion_update}
\end{equation}
in which $\odot$ denotes quaternion multiplication. The resulting
quaternion is normalized after each update to suppress numerical drift.
For vanishing angular velocity, $\Delta\mathbf{q}^{\,n}$ is taken to be
the identity quaternion. The rotation matrix
$\V{\mathcal{R}}^{\,n+1}$ associated with
$\overline{\mathbf{q}}^{\,n+1}$ then maps body-fixed vectors to the
laboratory frame. Thus, the mean orientation of a material vector $\R$
is obtained as
$\overline{\r}^{n+1}=\V{\mathcal{R}}^{\,n+1}\R$. The updated center-of-mass position and
orientation are subsequently used to reconstruct the immersed-body
indicator for the next acoustic solve.


\subsection{Time marching algorithm}
\label{subsec:time_marching}

\begin{figure}[t]
    \centering
    \includegraphics[width=0.9\linewidth]{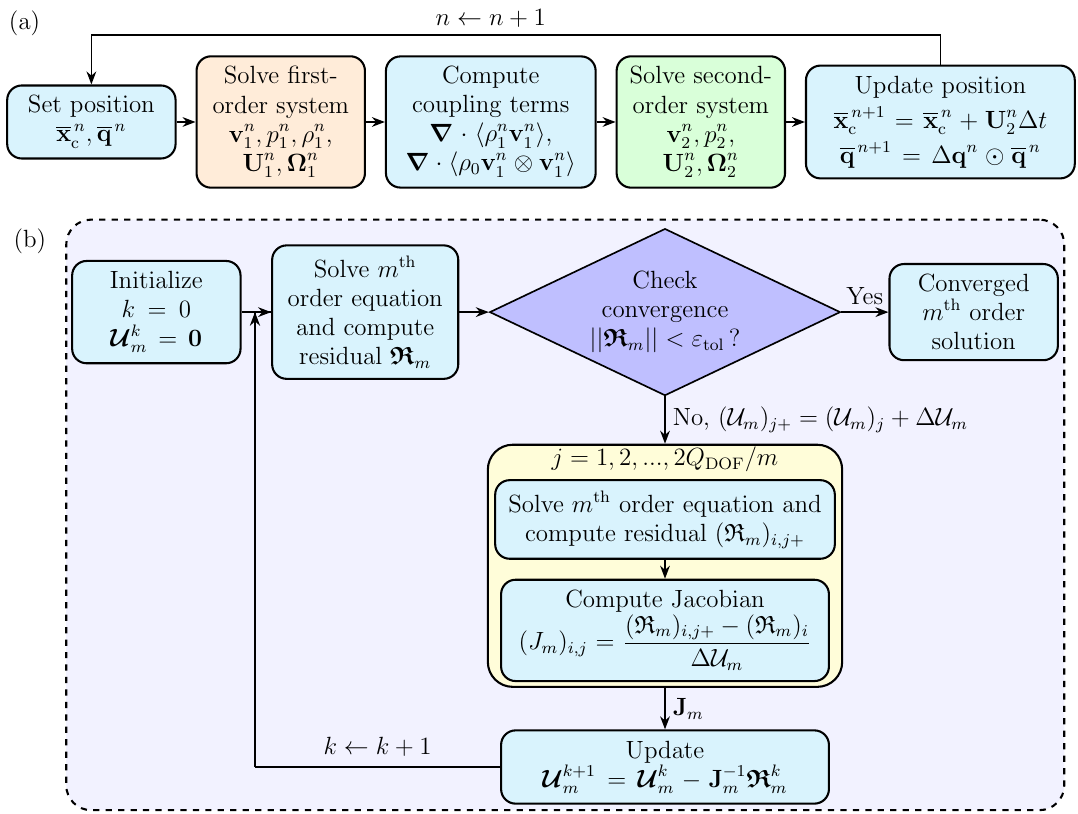}
    \caption{Flowchart illustrating the algorithm used to solve for and update the rigid body's position and orientation. (a) shows the overall time loop with update of position and orientation of the rigid body. (b) shows the iterative algorithm to find the converged first- and second-order solutions satisfying the rigid body motion constraints with Newton--Raphson method.}
    \label{fig:flowchart_time_marching}
\end{figure}

The numerical procedure used to advance the equations of motion over one
time step is summarized in Fig.~\ref{fig:flowchart_time_marching}.
Fig.~\ref{fig:flowchart_time_marching}(a) illustrates the overall
procedure for updating the position and orientation of the rigid body.
At time-step index $n$, the current mean center-of-mass position
$\overline{\x}^{\,n}_{\rm c}$ and orientation $\overline{\V{q}}^{\,n}$ are used
to solve the first-order problem
(Eq.~\eqref{eq:matrix-first-brinkman}) for the oscillatory flow fields
$(\vone^{\,n},\pone^{\,n},\rhoone^{\,n})$ and the corresponding
rigid-body velocities $(\U_1^{\,n},\bm{\Omega}_1^{\,n})$. The resulting
first-order fields are then used to evaluate the time-averaged nonlinear
coupling terms entering the second-order problem
(Eq.~\eqref{eq:matrix-second-brinkman}). The second-order problem is
subsequently solved for the mean flow fields
$(\ptwo^{\,n},\vtwo^{\,n})$ and rigid-body velocities
$(\U_2^{\,n},\bm{\Omega}_2^{\,n})$. These quantities are then used to
advance the mean center-of-mass position
$\overline{\x}_{\rm c}^{\,n+1}$ and orientation
$\overline{\V{q}}^{\,n+1}$ according to
Eqs.~\eqref{eq:update_xc_eq} and~\eqref{eq:mean_quaternion_update},
respectively. The time-step size $\Delta t$ is selected based on the
analysis presented in Sec.~\ref{sec:timestepanalysis}.

Both the first- and second-order problems are solved iteratively using the
Newton--Raphson procedure summarized in
Fig.~\ref{fig:flowchart_time_marching}(b). For a given perturbation order
$m$, with $m=1$ and $m=2$ corresponding to the first- and second-order
systems, respectively, the unknown rigid-body velocity vector
$\boldsymbol{\mathcal{U}}_m=\{\U_m,\bm{\Omega}_m\}$ is initialized to zero.
The $m^{\rm th}$-order equations are then solved, the corresponding
hydrodynamic force and torque $(\fhydro_m,\thydro_m)$ are evaluated, and
the residual vector $\bm{\mathfrak{R}}_m$ is constructed.

At first order ($m=1$), the residual follows from the rigid-body constraint
in Eq.~\eqref{eq:FO_constraint_real_imag} and is given by
\begin{align}
\label{eq:FO_residual}
    \begin{bmatrix}
        \omega \rhop\volume_\mathrm{p} & 0 & 0 & 0 \\
        0 & -\omega \rhop\volume_\mathrm{p} & 0 & 0 \\
        0 & 0 & \omega \bm{\mathcal{I}}_c^{\rm lab} & 0 \\
        0 & 0 & 0 & -\omega \bm{\mathcal{I}}_c^{\rm lab}
    \end{bmatrix}
    \begin{bmatrix}
        \U_1^{\rm r} \\
        \U_1^{\rm i} \\
        \bm{\Omega}_1^{\rm r} \\
        \bm{\Omega}_1^{\rm i}
    \end{bmatrix}
    -
    \begin{bmatrix}
        \fhydro_1^{\,\rm i} \\
        \fhydro_1^{\,\rm r} \\
        \thydro_1^{\,\rm i} \\
        \thydro_1^{\,\rm r}
    \end{bmatrix}
    =
    \bm{\mathfrak{R}}_1 .
\end{align}
At second order ($m=2$), the quasi-steady force- and torque-free
constraints give
\begin{equation}
    \bm{\mathfrak{R}}_2
    =
    \begin{bmatrix}
        \langle \fhydro_2\rangle \\
        \langle \thydro_2\rangle
    \end{bmatrix}.
\end{equation}
For each perturbation order $m$, convergence is declared when the norm of
the residual $\bm{\mathfrak{R}}_m$ satisfies a tolerance of
$\varepsilon_{\mathrm{tol}}=10^{-15}$. If the convergence criterion is not satisfied, each
component $(\mathcal{U}_{m})_{j}$ of the unknown vector
$\boldsymbol{\mathcal{U}}_m$ is perturbed individually according to
\begin{equation}
    (\mathcal{U}_{m})_{j+}
    =
    (\mathcal{U}_{m})_{j}
    +
    \Delta\mathcal{U}_m .
\end{equation}
The resulting residual $(\mathfrak{R}_{m})_{i,j+}$ is used to approximate the
corresponding entry of the Jacobian matrix $\mathbf{J}_m$ by
\begin{equation}
    (J_m)_{i,j}
    =
    \frac{
        (\mathfrak{R}_{m})_{i,j+}
        -
        (\mathfrak{R}_{m})_{i}
    }{
        \Delta\mathcal{U}_m
    } .
\end{equation}
For the $m^{\rm th}$-order problem, the indices $i$ and $j$ range from
$1$ to $2Q_{\rm DOF}/m$, where $Q_{\rm DOF}$ denotes the number of free
rigid-body degrees of freedom. The maximum value of $Q_{\rm DOF}$ is $6$ in three dimensions
and $3$ in two dimensions, and the corresponding Jacobian $\V{J}_m$
has dimensions $(2Q_{\rm DOF}/m)\times(2Q_{\rm DOF}/m)$. In the present
work, all three rigid-body degrees of freedom are unconstrained in the
two-dimensional computational domain, resulting in Jacobian matrices
$\mathbf{J}_1$ and $\mathbf{J}_2$ of size $6\times6$ and $3\times3$,
respectively.

At Newton iteration $k$, the unknown vector is updated according to
\begin{equation}
    \boldsymbol{\mathcal{U}}_m^{k+1}
    =
    \boldsymbol{\mathcal{U}}_m^{k}
    -
    \mathbf{J}_m^{-1}
    \bm{\mathfrak{R}}_m^{k},
\end{equation}
and the procedure is repeated until the convergence criterion is
satisfied. We use perturbation magnitudes
$\Delta\mathcal{U}_1=10^{-2}$ and
$\Delta\mathcal{U}_2=10^{-4}$ for the first- and second-order problems,
respectively. These values yield convergence within at most two Newton
iterations for the cases considered here.

\subsection{Linear solvers}
The first-order system in Eq.~\eqref{eq:matrix-first-brinkman} consists
of coupled Helmholtz equations for the complex amplitudes $\vonehat$ and
$\ponehat$. Owing to the challenges associated with efficiently solving
the coupled Helmholtz system iteratively, we employ the sparse direct
solver MUMPS~\cite{MUMPS:1}. The second-order system in
Eq.~\eqref{eq:matrix-second-brinkman} is a steady, low-Mach-number,
Brinkman-penalized Stokes system. We solve this system using FGMRES with
a projection-based preconditioner that incorporates the Brinkman
penalization term. Further details of the preconditioner and its
implementation are provided in our previous work~\cite{gautam2025simulating}.

\section{Software implementation}

The perturbation-based fluid--structure interaction solver
developed in this work is implemented within IBAMR
\cite{IBAMR-web-page}, an open-source C++ library for
computational fluid dynamics and immersed boundary method
development. The IBAMR source code is publicly available
at \url{https://github.com/IBAMR/IBAMR}.
IBAMR employs the SAMRAI library
\cite{HornungKohn02, samrai-web-page} for Cartesian grid
management and the PETSc library
\cite{petsc-user-ref, petsc-web-page} for numerical
linear algebra and solver infrastructure.
     
\section{Results}
In this section, we present a series of representative test cases to assess the efficacy of the proposed framework for simulating the dynamics of immersed rigid bodies in acoustofluidic systems. We first consider a circular particle undergoing purely translational motion and examine the sensitivity of the predicted dynamics to the time-step size. We then investigate the influence of particle density and size on the translational response. Finally, we consider an elliptical particle to demonstrate the capability of the framework to capture coupled translational and rotational motion in acoustically actuated microchannels.


\subsection{Rigid body translation}
\subsubsection{Dynamics}
\label{sec:translation_only}
\begin{figure}
    \centering
    \includegraphics[width=0.75\linewidth]{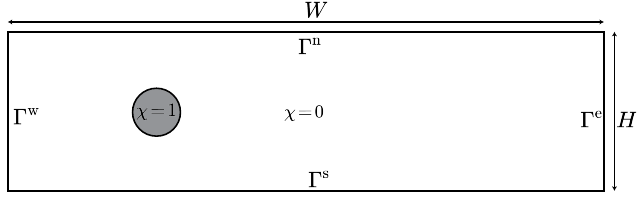}
    \caption{Schematic of a fluid-filled, rectangular microchannel comprising a rigid circle, denoted in grey. The channel dimensions are taken as $W = \SI{150}{\micro\metre}$ and $H= \SI{40}{\micro\metre}$. The immersed domain is circular with radius  $R_0 = \SI{5}{\micro\metre}$, and its center is located at $(W/4, H/2)$. The device is acoustically actuated by prescribing a rectilinear actuation at the left and right walls ($\Gamma^\textrm{w} \cup \Gamma^\textrm{e}$), while the top and bottom walls ($\Gamma^\textrm{n} \cup \Gamma^\textrm{s}$) are taken to be fixed.}
    \label{fig:schematic}
\end{figure}

To demonstrate the purely translational motion of the rigid body under acoustic forcing, we consider a circular particle in a rectangular microchannel of width $W = \SI{150}{\micro\metre}$ and height $H = \SI{40}{\micro\metre}$ (see Fig.~\ref{fig:schematic}). The channel is occupied by water modeled as a compressible Newtonian fluid with density $\rhozero = 998~\mathrm{kg/m}^3$, shear viscosity $\mu = \SI{0.89}{\textrm{mPa$\cdot$\second}}$, second viscosity $\lambda = \SI{1.88}{\textrm{mPa$\cdot$\second}}$, and speed of sound $c_0 = \SI{1500}{\meter/\second}$. The immersed object has a radius of $R_0 = \SI{5}{\micro\metre}$, with its center initially located at $(x_0, y_0) = (W/4, H/2)$. For this specific case, we consider the density of the immersed solid to be the same as water.


The immersed solid region, centered at the location $(\overline{x}_{\rm c}, \overline{y}_{\rm c})$, is identified via a signed distance function $\phi(\x)$ defined as
\begin{equation}\label{eq:circle_sdf}
\phi(\x) = \sqrt{(x-\overline{x}_{\rm c})^2 + (y-\overline{y}_{\rm c})^2}- R_0,
\end{equation}
such that the zero level set of $\phi(\x)$ denotes the fluid--solid interface. 

At the first order, rectilinear oscillations of the left and right walls ($\Gamma^\mathrm{w}$ and $\Gamma^\mathrm{e}$) are prescribed via velocity boundary conditions $\v_1 = \{\omega \d_a e^{i\omega t},0\}$, where $\d_a = \SI{0.1}{\nano\meter}$ is the displacement amplitude and $\omega = 2\pi f$ is the angular frequency. The system is driven at an acoustic frequency $f = \SI{5}{MHz}$, corresponding to the one-dimensional half-wave resonance condition $f = c_0/(2W)$~\cite{bruus2012acoustofluidics}. This actuation establishes a pressure node at the center of the channel $x=W/2$, analogous to the experimental setup previously employed by Muller \etal~\cite{muller2013ultrasound}.  No-slip velocity boundary conditions are enforced on the top and bottom walls ($\Gamma^\textrm{n}\,\cup\,\Gamma^\textrm{s}$).
Further, the rigid solid is modeled via a volume penalization approach by enforcing the first-order rigid body constraint, as described in Sec.~\ref{subsec:time_marching}. At second order, all the surrounding walls are prescribed a zero mean Lagrangian velocity boundary condition, $\langle \v_2\rangle  = -\v^{\textrm{SD}}$. The rigid solid is again modeled by prescribing the second-order rigid body constraint, as described in Sec.~\ref{subsec:time_marching}. The computational domain is discretized using a uniform Cartesian mesh with $(N_x\times N_y) = (3000\times1200)$ grid cells. To enforce the constraints via the volume penalization approach, we consider a penalty factor $p_\perm=10^{10}$ and smearing width $n_\textrm{cells}=1$, which were found to be adequate in our prior work for a similar setup~\cite{gautam2025simulating}.

\begin{figure}
    \centering
    \includegraphics[width=\linewidth]{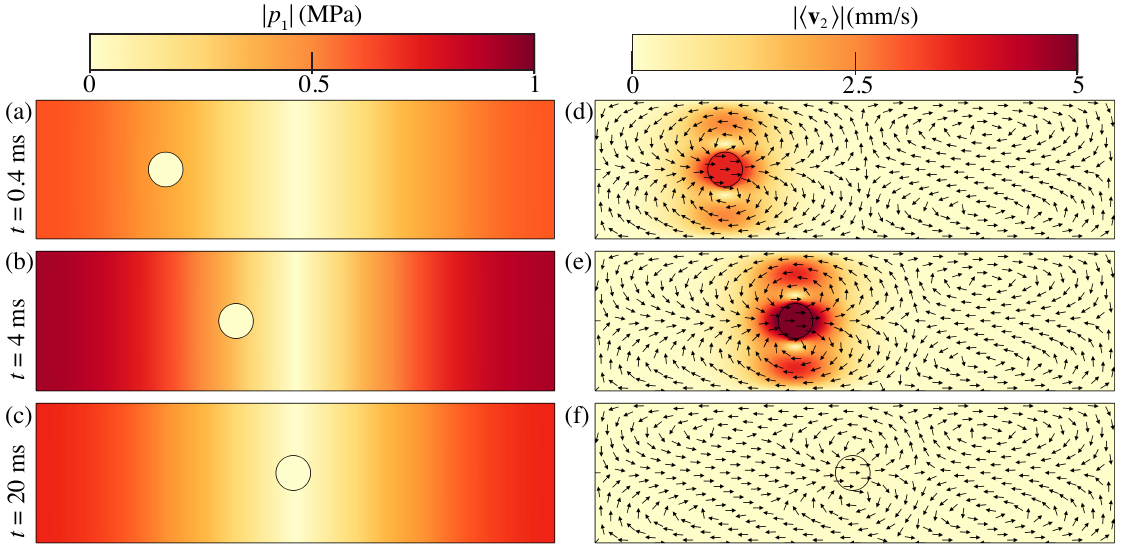}
    \caption{Position of a circular rigid object at three different time instants: (a,d) $t=0.4$ ms, (b,e) $t=4$ ms, and (c,f) $t=20$ ms.  Color in the left panels (a--c) indicates the pressure amplitude defined as $|p_1|=\sqrt{(p_1^{\rm r})^2+(p_1^{\rm i})^2}$. Color in the right panels (d--f) indicates the magnitude of the time-averaged, second-order velocity, while arrows denote the corresponding normalized velocity vectors.}
    \label{fig:translation_motion}
\end{figure}

Prior to discussing the numerical results, we introduce the acoustic contrast factor ($\Phi$), which characterizes the relative acoustic properties of the particle and the surrounding fluid. For a spherical particle, it is defined as
\begin{subequations}\label{eq:ACF}
\begin{alignat}{2}
\Phi =& \frac{1}{3} \Tilde{f}_1 + \frac{1}{2}\Tilde{f}_2, \\ 
\quad \Tilde{f}_1 =& 1-\frac{\kappa_{\mathrm{p}}}{\kappa_0},\quad\Tilde{f}_2 = \operatorname{Re}\left [\frac{2(\Tilde{\rho}-1)(1-\gamma)}{2\Tilde{\rho}+1-3\gamma}  \right],\\
\quad \Tilde{\rho} = \frac{\rhop}{\rhozero},\quad \gamma  =& -\frac{3}{2}\left[1-\iota (1+\Tilde{\delta})\right]\Tilde{\delta},\quad \Tilde{\delta} = \frac{\delta}{R_0},\quad\delta = \sqrt{\frac{2\mu}{\omega \rhozero}},
\end{alignat}
\end{subequations}
in which $\Tilde{f}_1$ and $\Tilde{f}_2$ are the monopole and dipole scattering coefficients, respectively, $\delta$ is the viscous boundary layer thickness,  and $\kappa_{\rm p}$ and $\kappa_{0}$ denote the particle and fluid compressibilities, respectively~\cite{settnes2012forces}. In a one-dimensional standing acoustic wave, the sign of the acoustic contrast factor determines whether a particle migrates toward a pressure node or antinode. Particles with $\Phi>0$ migrate toward pressure nodes, whereas those with $\Phi<0$ migrate toward pressure antinodes. The acoustic contrast factor therefore provides a convenient indicator of the direction of particle migration induced by the acoustic radiation force. In the present work, we consider rigid particles, for which $\kappa_{\rm p}=0$, so that the acoustic contrast factor depends only on the density ratio $\Tilde{\rho}$ and the normalized boundary-layer thickness $\Tilde{\delta}$. For $\rhop=\rhozero=998~\mathrm{kg/m}^3$, Eq.~\eqref{eq:ACF} gives $\Phi=1/3$, corresponding to migration toward the pressure node. Although Eq.~\eqref{eq:ACF} is derived for a spherical particle, we use it here only as a qualitative indicator of the migration direction for the two-dimensional (cylindrical) particles considered in this work.

Figs.~\ref{fig:translation_motion}(a,b,c) show the numerically predicted mean position of the immersed particle at different times, with the color indicating the pressure amplitude. As expected, the particle migrates toward the pressure node located at the channel centerline, $x=W/2$. Owing to its symmetric placement at $y=H/2$, the particle does not undergo rotational motion. Figs.~\ref{fig:translation_motion}(d,e,f) show the corresponding second-order streaming velocity fields. At $t=0.4~\mathrm{ms}$, the flow in each half of the channel is characterized by a pair of counter-rotating vortices, with the pair in the left half surrounding the particle. The resulting streaming pattern is qualitatively consistent with experimentally observed flow fields in similar configurations without an immersed particle~\cite{muller2013ultrasound}. In the absence of the particle, the streaming field is symmetric about the channel centerline. The immersed particle breaks this symmetry, however, leading to substantially larger streaming velocities in the left half of the channel.
%
%
%
%
As the particle moves toward the pressure node (Figs.~\ref{fig:translation_motion}(e,f)), its influence progressively extends into the right half of the channel. In particular, the vortices in the right half of the channel become increasingly compressed in the horizontal direction. At all time instants, the maximum flow velocity occurs near the fluid--solid interface and is significantly larger than the streaming velocity in the absence of the particle.

\begin{figure}
    \centering
    \includegraphics[width=\linewidth]{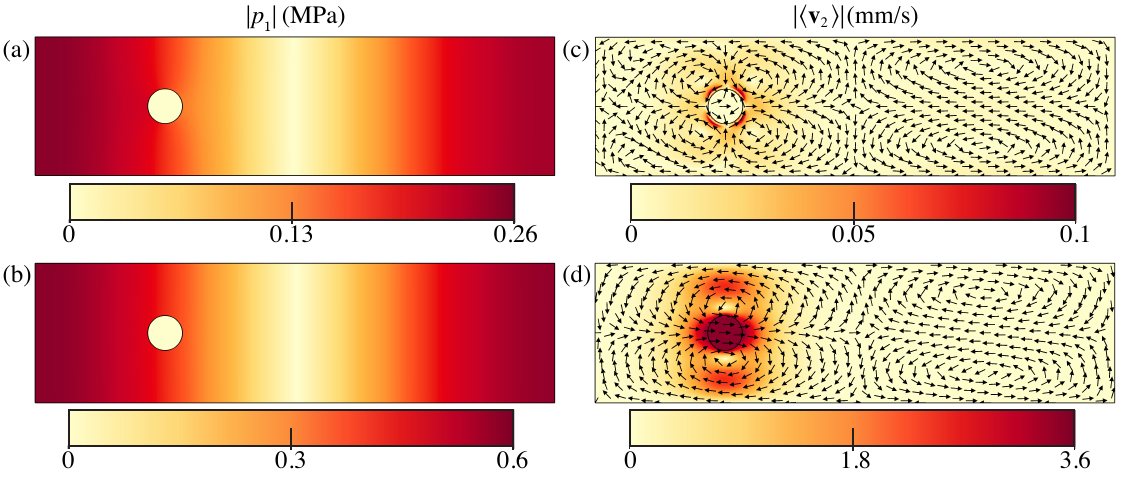}
    \caption{Comparison of the first- and second-order solution for (a,c) a fixed particle and (b,d) a moving particle. (a,b) plot the pressure amplitude defined as $|p_1|=\sqrt{(p_1^{\rm r})^2+(p_1^{\rm i})^2}$. (c,d) plot the magnitude of the time-averaged, second-order velocity with arrows denoting the corresponding normalized velocity vectors.}
    \label{fig:pinned_moving}
\end{figure}
Next, we perform simulations for the same configuration while holding the fluid--solid interface fixed at both first and second order. This is achieved by prescribing zero structural velocities at each order, $\vpenalty_1=\vpenalty_2=\bm{0}$. This comparison is motivated by a common approach in the acoustofluidics literature: in the absence of a framework for resolving particle motion, the flow field is computed around a fixed particle and compared with experimentally observed flow patterns to infer particle oscillation modes that are otherwise difficult to measure or visualize directly. Comparing the fixed-particle and moving-particle cases therefore provides a means of assessing the validity and limitations of this commonly adopted approach.

%
%
Fig.~\ref{fig:pinned_moving} compares the simulation results obtained for a fixed immersed particle with those obtained for a freely moving immersed particle. Comparing the first-order pressure fields in Figs.~\ref{fig:pinned_moving}(a,b), a half-wave standing field is observed in both cases, although the maximum pressure amplitudes differ, with values of $\SI{0.26}{MPa}$ and $\SI{0.6}{MPa}$ for the fixed and moving particles, respectively. Despite the qualitative similarity in the first-order response, the corresponding second-order streaming velocity fields are markedly different (Figs.~\ref{fig:pinned_moving}(c,d)) under otherwise identical conditions. The flow around the fixed particle is characterized by four vortices surrounding the particle, which is the classical streaming pattern generated by the relative oscillation between a cylinder and the surrounding fluid along a single direction, here the $x$ direction~\cite{longuet-higgins_viscous_1998}. In contrast, the flow around the freely moving particle exhibits two vortices. Although the acoustic field drives the fluid along the $x$ direction in both cases, the freely moving particle oscillates together with the surrounding fluid, which substantially reduces the relative oscillation between the two and hence the corresponding four-vortex contribution. Instead, the streaming is dominated by fluid dragged along by the particle as it migrates toward the pressure node, which produces the two-vortex pattern. The streaming pattern therefore reflects the particle's freedom to move rather than a different direction of oscillation and should not, in general, be used to infer the particle's oscillation mode without accounting for particle mobility.



The underlying difference can be traced to the interfacial velocity. For the fixed particle, both the first- and second-order velocities vanish at the fluid--solid interface, and the corresponding streaming velocity is therefore zero at the particle surface. However, for the freely moving particle, the no-slip condition requires the fluid to move with the particle, and the second-order velocity at the interface is set by the mean particle velocity and the Stokes drift associated with the particle oscillation (Eq.~\eqref{eq:v2b_mean_final}), and is therefore nonzero.
This distinction also produces a substantial difference in the streaming magnitude: the maximum streaming velocity for the freely moving particle is more than an order of magnitude larger than that for the fixed particle. Moreover, for the moving particle, the maximum occurs directly at the fluid--solid interface, whereas for the fixed particle it is displaced slightly into the fluid, within the boundary layer. These results demonstrate that particle mobility can fundamentally alter not only the topology of the acoustic streaming field, but also its magnitude and spatial distribution.
%
%
In addition, the present framework differs from prior approaches~\cite{voss2022acoustic} in how the particle velocity is obtained. In those approaches, the mean particle velocity is typically determined from a force balance in which the acoustic radiation force is equated to a hydrodynamic drag force, with the drag evaluated through a hydrodynamic resistance matrix computed separately for the particle. In the present framework, the mean translational and rotational velocities of the particle are instead obtained directly as part of the solution by enforcing the force- and torque-free conditions on the coupled fluid--particle system. Consequently, no separate drag model or resistance matrix is required, and the particle velocity and trajectory can be computed for particles of general shape with full two-way coupling.

\subsubsection{Selection of time-step size: accuracy vs. efficiency}
\label{sec:timestepanalysis}

In this section, we investigate the impact of the choice of time step size on the accuracy of the predicted particle dynamics.
Referring to Sec.~\ref{subsec:time_marching}, the choice of time-step size does not directly affect the solvability of the numerical system of equations since the problem is split into a time-harmonic first-order problem and a steady second-order problem. As such, time enters the system of equations only during the update of the particle position and orientation, and therefore affects only the accuracy of the solution, rather than the solvability of the problem itself.


\begin{figure}
    \centering
    \includegraphics[width=0.5\linewidth]{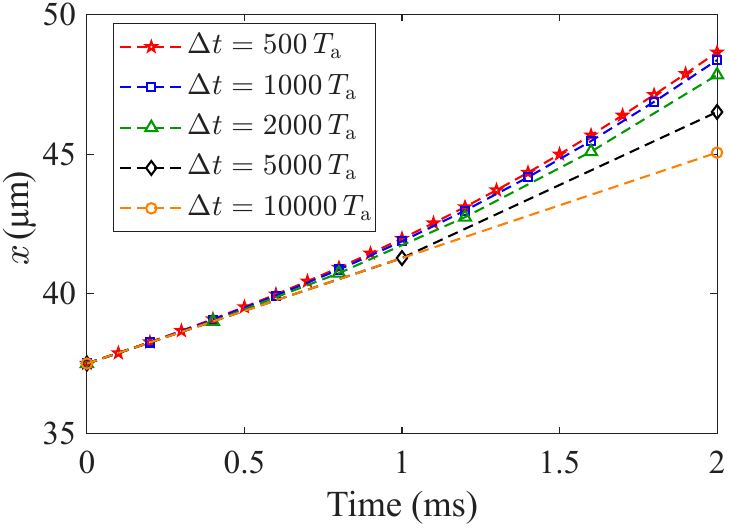}
    \caption{Predicted particle position along the $x$ direction as a function of time obtained from simulations with different choices of time step size.}
    \label{fig:time_step}
\end{figure}

To assess the effect of the time-step size, we consider the same configuration as described in Sec.~\ref{sec:translation_only}. The rigid particle is initially located at $(x_0, y_0) = (W/4, H/2)$, and its trajectory is computed using time-step sizes ranging from $\Delta t=500\,\tp$ to $10{,}000\,\tp$. In each simulation, the particle motion is tracked up to a physical time of $t=2~\mathrm{ms}$. Fig.~\ref{fig:time_step} shows the particle position in the $x$ direction as a function of time for the different time-step sizes. As the time-step size is decreased, the predicted trajectories progressively converge, indicating that the particle dynamics remain well resolved even for relatively large time steps. 

These results also illustrate the computational advantage of the present approach relative to a direct simulation that resolves the acoustic time scale. For example, advancing the solution by a single time step of $\Delta t=500\,\tp$ would require approximately $5000$ acoustic-resolving time steps if ten time steps are used to resolve each acoustic period. The proposed framework therefore captures the translational dynamics of the immersed rigid particle using time steps that are orders of magnitude larger than those required to directly resolve the underlying acoustic oscillations, substantially reducing the computational cost of long-time simulations of acoustically driven particle motion. We note that the time-step size used to assess the accuracy of the solution in this case should not be considered generally applicable to other cases. A time-step convergence study should be performed for each case with different operating parameters to select the appropriate time step size, as is done for the other cases presented in this work.

\subsection{Effect of density of the particles on translation}
\label{sec:density}
\begin{figure}
    \centering
    \includegraphics[width=\linewidth]{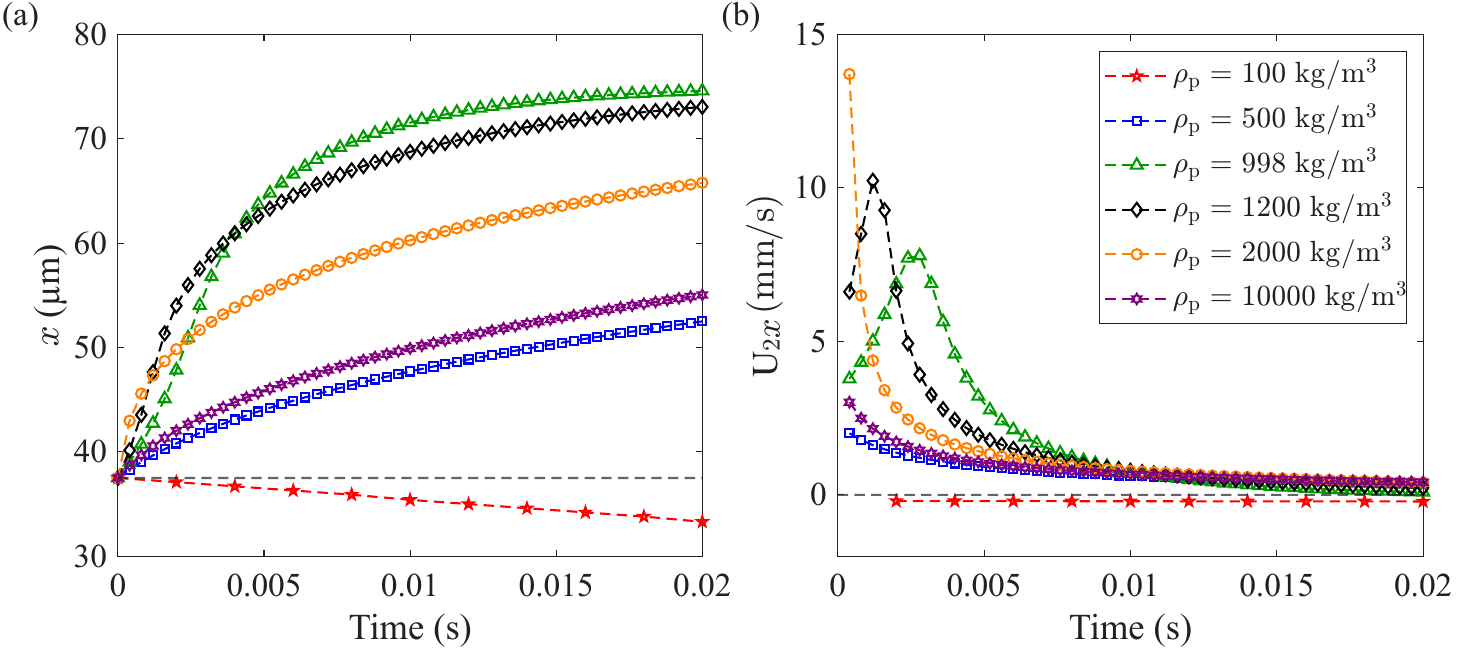}
    \caption{(a) Position of the immersed particle along the $x$  direction as a function of time for varying particle density. (b) Translational velocity of the particle ($x$-component) as a function of time.}
    \label{fig:density_plot}
\end{figure}



Next, we investigate the influence of particle density on its motion. Classically, the acoustic radiation force is characterized in terms of the acoustic contrast factor, which depends on the relative acoustic properties of the particle and the surrounding fluid (see Eq.~\eqref{eq:ACF}). Particles with different densities are therefore expected to exhibit distinct mean dynamics, a dependence that has been exploited experimentally for density-based particle sorting~\cite{ma2016mechanical,xie2020acoustic}.

We consider the same configuration as in Sec.~\ref{sec:translation_only} and vary the rigid-particle density as $\rhop=$ $100$, $500$, $998$, $1200$, $2000$, and $10,000\,\mathrm{kg/m^3}$. The corresponding acoustic contrast factors are $\Phi=$ -0.34, 0.09, 0.33, 0.39, 0.54, and 0.79, respectively. Fig.~\ref{fig:density_plot}(a) shows the particle $x$-position as a function of time for the different densities.
The particle with $\rhop=100\,\mathrm{kg/m^3}$ migrates toward the pressure antinode located at $x=0$, consistent with its negative acoustic contrast factor. All other particles migrate toward the pressure node located at the channel centerline, $x=W/2$, consistent with their positive acoustic contrast factors. Among the cases considered, the neutrally buoyant particle reaches the pressure node in the shortest time. For particles denser than the surrounding fluid, increasing the particle density results in progressively slower migration toward the pressure node.

These results are noteworthy in the context of the conventional interpretation of acoustic-radiation-force-driven particle migration. A commonly adopted reduced approach uses first- and second-order fields obtained from a particle-free simulation to predict particle trajectories. In this approach, the acoustic and streaming fields are first computed by solving a fluid-only problem without explicitly including the immersed particle. The mean hydrodynamic force, $\bm\fhydro_2$, acting on a spherical particle is then modeled as the sum of (a) an acoustic radiation force arising from scattering of the acoustic field by the particle, $\V{F^\textrm{rad}}=f(\pone,\vone,\Phi)$, and (b) a viscous drag force associated with the acoustic streaming flow, $\V{F^\textrm{d}}=-6\pi R_0\mu(\v_{\rm p}-\vtwo)$, in which $\v_{\rm p}$ denotes the particle velocity. Neglecting particle inertia, the total mean hydrodynamic force is set to zero, yielding
$\v_{\rm p}=\vtwo+\frac{\V{F^\textrm{rad}}}{6\pi R_0\mu}$.
For a horizontal pressure half-wave resonance, $\V{F^\textrm{rad}}\;\propto\;\Phi$, suggesting that particles with progressively larger contrast factors should experience stronger radiation forces.

The results in Fig.~\ref{fig:density_plot}(a), however, exhibit a different trend. For particles denser than the surrounding fluid, migration toward the pressure node becomes slower as the density increases. Furthermore, the particle with $\rhop=100\,\mathrm{kg/m^3}$ migrates toward the pressure antinode more slowly than the particles with $\rhop=500$ and $998\,\mathrm{kg/m^3}$, despite having a comparable or larger $|\Phi|$. These differences arise because the reduced approach assumes that the immersed particle experiences the same first- and second-order fields as those obtained in its absence, thereby neglecting the two-way coupling between the particle and the surrounding fluid. In the present formulation, by contrast, the particle density directly influences the first-order dynamics through the rigid-body equations of motion (see Eqs.~\eqref{eq:FO_Newtons_balance}). The resulting changes in the first-order particle response modify both the forcing and boundary conditions of the second-order problem, leading to distinct mean particle dynamics. 

Fig.~\ref{fig:density_plot}(b) shows the translational particle velocity as a function of time. Referring to Eq.~\eqref{eq:SO_Newtons_balance_constraint}, the mean particle dynamics are treated as quasi-steady: at each instant, the mean hydrodynamic force on the particle vanishes, and the particle therefore assumes the velocity required to remain force-free at its instantaneous position. This behavior is analogous to particle motion in steady Stokes flow, for which the instantaneous particle velocity is determined by the local force balance rather than by its acceleration history. Consequently, the particle velocity in Fig.~\ref{fig:density_plot}(b) varies in time as the particle migrates through the spatially nonuniform acoustic field. The radiation and streaming contributions vary with position, and the velocity required to maintain zero net force therefore changes along the particle trajectory. Accordingly, the temporal evolution shown in Fig.~\ref{fig:density_plot}(b) also reflects the spatial variation of the particle speed throughout the microchannel.

Particles of different densities attain their maximum velocities at different times and spatial locations. More importantly, Fig.~\ref{fig:density_plot}(b) shows that the ordering of particle velocities does not, in general, follow that implied solely by the acoustic contrast factor. Within the reduced description, the dependence $\V{F^\textrm{rad}} \;\propto\;\Phi$ suggests that particles with larger contrast factors should exhibit stronger radiation-force-driven migration. The present simulations show that this ordering is not generally preserved: depending on its instantaneous position in the channel, a denser particle may exhibit either a smaller or larger velocity than a lighter particle.

\begin{figure}
    \centering
    \includegraphics[width=\linewidth]{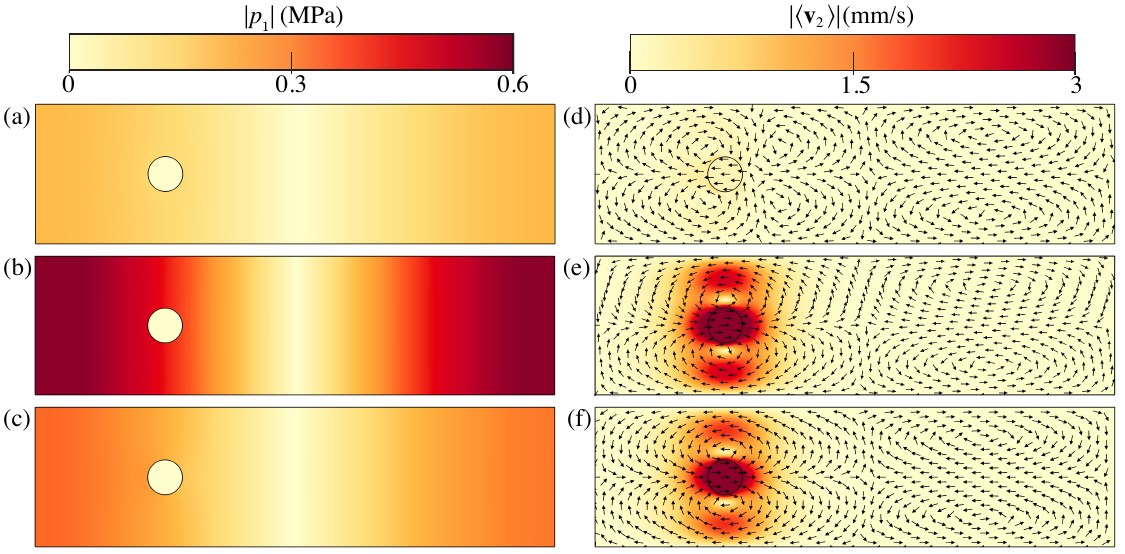}
    \caption{Comparison of the first- and second-order solution for three different particle densities: (a,d) $\rhop=100\,\mathrm{kg/m^3}$, (b,e) $\rhop=998\,\mathrm{kg/m^3}$, and (c,f) $\rhop=10,000\,\mathrm{kg/m^3}$. (a--c) plot the pressure amplitude defined as $|p_1|=\sqrt{(p_1^{\rm r})^2+(p_1^{\rm i})^2}$. (d--f) plot the magnitude of the time-averaged, second-order velocity with arrows denoting the corresponding normalized velocity vectors.}
    \label{fig:Density_variation}
\end{figure}

Figs.~\ref{fig:Density_variation}(a--c) show the first-order pressure fields for three particle densities, $\rhop=$ 100, 998, and 10,000 $\mathrm{kg/m^3}$. Although a standing-wave pattern is obtained in all three cases, the pressure amplitude varies substantially with particle density and is largest for the neutrally buoyant case, $\rhop=$ 998 $\mathrm{kg/m^3}$. The corresponding second-order velocity fields are shown in Figs.~\ref{fig:Density_variation}(d--f). For $\rhop=$100 $\mathrm{kg/m^3}$, the streaming topology differs qualitatively from those of the other two cases and contains an additional pair of background streaming vortices in the left half of the channel. At the location considered $(x_0, y_0) = (W/4, H/2)$, the largest streaming and particle velocities are observed for the neutrally buoyant particle, consistent with the largest first-order pressure amplitude in this case.


Overall, the results in Figs.~\ref{fig:density_plot} and~\ref{fig:Density_variation} demonstrate that particle density affects the coupled dynamics in a manner that is not captured solely by the acoustic contrast factor. These findings further underscore the importance of explicitly resolving fluid--structure interactions by incorporating the moving particle directly into the formulation, rather than inferring particle dynamics through post-processing of particle-free simulations.

\subsection{Translation of a small particle}




\begin{figure}
    \centering
    \includegraphics[width=0.75\linewidth]{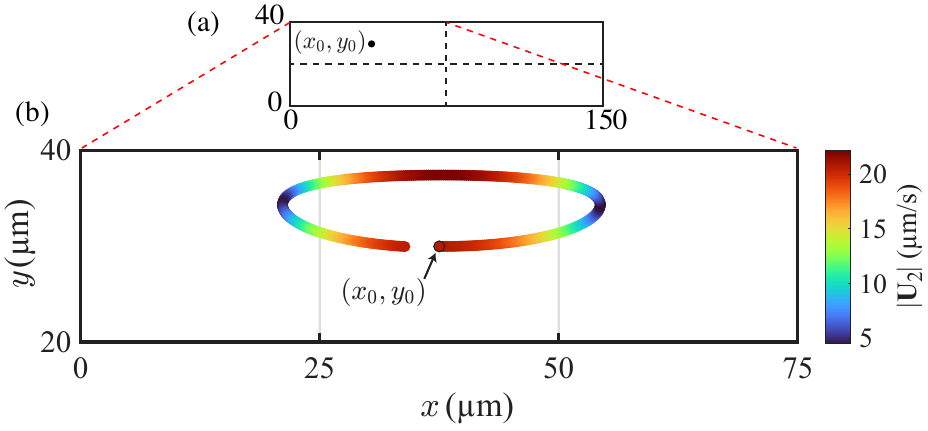}
    \caption{Trajectory of a particle of radius $R_0=\SI{0.1}{\micro\meter}$, with the bead color representing the magnitude of the particle's translational velocity $|\mathbf{U}_2|$. The particle is seeded at the initial location $(x_0,y_0) = (W/4,3H/4)$, where $W$ and $H$ denote the width and height of the channel, respectively. The bead size is not proportional to the particle size.}
    \label{fig:tracer}
\end{figure}

In the test cases considered thus far, the particle migrates toward or away from the pressure node depending on the sign of its acoustic contrast factor. Owing to the geometric symmetry of these configurations, with a circular particle initially positioned along the horizontal midline of the channel, the resulting motion is rectilinear. We next consider a regime in which the particle dynamics are dominated by the underlying fluid motion. This test case is motivated by the use of small tracer particles in acoustofluidic experiments, where their motion is commonly used to visualize and quantify acoustic streaming.
We consider a particle of radius $R_0 = \SI{0.1}{\micro\metre}$, initially located at $(x_0,y_0) = (W/4,3H/4)$, in which $W$ and $H$ denote the channel width and height, respectively. A time-step size of $\Delta t = 5\times 10^4\,\tp$ was found to provide adequate temporal convergence for this case. The boundary conditions and all other operating parameters are identical to those used in Sec.~\ref{sec:translation_only}.

Fig.~\ref{fig:tracer} shows the particle trajectory over a total duration of $5$ s. As expected, the particle follows a curved path that closely resembles the acoustic streaming pattern obtained in the absence of the particle. However, whereas the corresponding streaming streamlines form closed vortices, the particle does not return to its initial position. 
The particle trajectory can be expected to approach the underlying streaming motion in the limit $R_0\rightarrow0$. Nevertheless, the close correspondence between the particle trajectory and the streaming pattern indicates that the particle motion is dominated by acoustic streaming, consistent with prior observations~\cite{muller2012numerical}.

Together with the results presented in Secs.~\ref{sec:translation_only} and~\ref{sec:density}, this test case demonstrates the capability of the proposed framework to capture two-dimensional immersed-particle dynamics across a range of particle sizes and densities, including regimes dominated by either acoustic radiation forces or acoustic streaming.

\subsection{Translation with rotation}
\label{sec:result:translation_rotation}


\begin{figure}
    \centering
    \includegraphics[width=\linewidth]{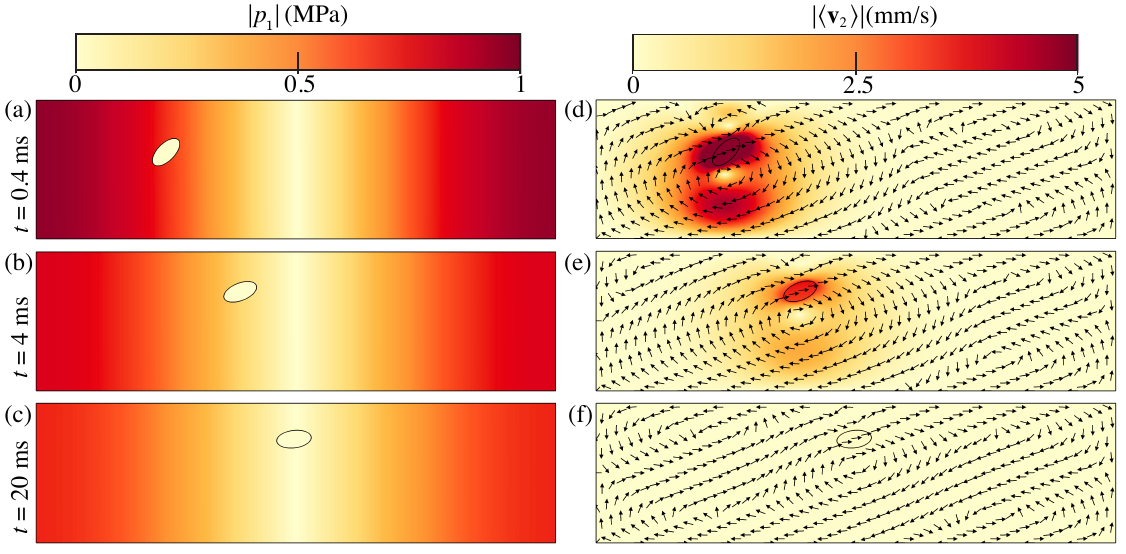}
    \caption{Position and orientation of an elliptical rigid object at three different time instants: (a,d) $t=0.4$ ms, (b,e) $t=4$ ms, and (c,f) $t=20$ ms.  Color in the left panels (a--c) indicates the pressure amplitude defined as $|p_1|=\sqrt{(p_1^{\rm r})^2+(p_1^{\rm i})^2}$. Color in the right panels (d--f) indicates the magnitude of the time-averaged, second-order velocity, while arrows denote the corresponding normalized velocity vectors.}
    \label{fig:rotation_SO}
\end{figure}

\begin{figure}
    \centering
    \includegraphics[width=\linewidth]{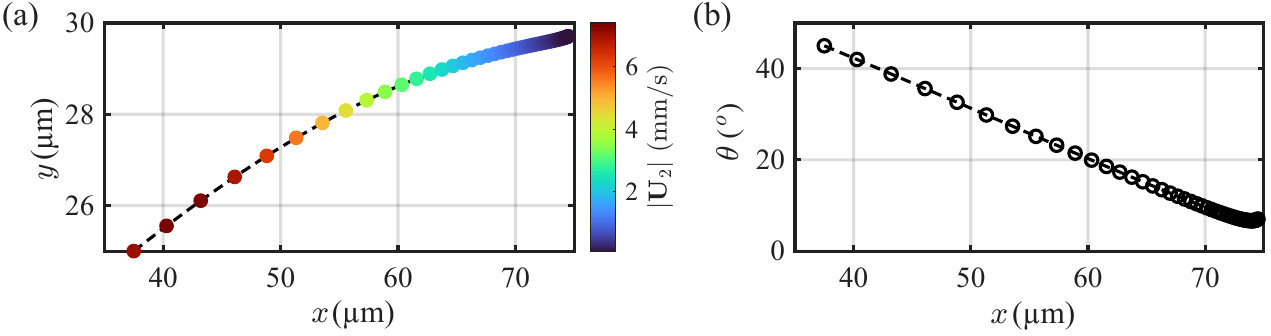}
    \caption{(a) Trajectory of an elliptical rigid particle as it migrates toward the pressure node. (b) Orientation ($\theta$) of the elliptical particle with respect to the horizontal axis as a function of its $x$-position. The total simulation time is 20 ms.}
    \label{fig:ellipse_trajectory}
\end{figure}

Having demonstrated purely translational particle motion, we now consider an elliptical particle to examine coupled translational and rotational dynamics. Unlike the preceding test cases, the elliptical particle experiences a nonzero first-order torque and attains a nonzero mean angular velocity. As discussed in Sec.~\ref{sec:governing_equations}, we formulate a consistent angular-momentum balance constraint at each perturbation order to update the particle orientation, analogous to the use of the linear-momentum balance constraint to update its position.
The channel dimensions, operating parameters, and fluid and particle properties are chosen to be identical to those of the translation-only case presented in Sec.~\ref{sec:translation_only}, unless otherwise stated. An ellipse with semi-major axis $a=\SI{5}{\micro\meter}$ and semi-minor axis $b = \SI{2.5}{\micro\meter}$ is initially positioned at $(x_0,y_0) = (W/4, 5H/8)$, with its major axis oriented at $45^\degree$ relative to the horizontal axis. The initial position and orientation are chosen to promote both translational and rotational motion. The time-step size is taken to be $\Delta t = 2000\,\tp$.

Figs.~\ref{fig:rotation_SO}(a--c) show the numerically obtained mean position and orientation of the particle at different time instants, with the color indicating the pressure amplitude. Similar to the circular-particle case shown in Fig.~\ref{fig:translation_motion}, the ellipse migrates from its initial position toward the pressure node. In addition, the particle rotates as it migrates, with its major axis gradually aligning perpendicular to the nodal plane. This motion is qualitatively consistent with the migration-phase behavior reported by Sachs et al.~\cite{sachs2023behavior} for prolate spheroids in a standing surface acoustic wave field.
Fig.~\ref{fig:rotation_SO} also shows the corresponding second-order streaming velocity field at different time instants. Because the particle is positioned away from the horizontal mid-axis of the channel, the resulting streaming field is asymmetric. This contrasts with the preceding circular-particle cases, in which the particle was positioned symmetrically at $y=H/2$, producing a streaming field symmetric about the horizontal mid-axis. The particle-induced streaming is observed to dominate the background streaming in the channel. The maximum streaming velocity occurs at the particle surface, where it is set by the particle’s mean surface velocity and the Stokes drift (Eq.~\eqref{eq:v2b_mean_final}).

Fig.~\ref{fig:ellipse_trajectory}(a) shows the particle trajectory in the $x$-$y$ plane. Unlike the circular particle, which undergoes rectilinear migration, the elliptical particle follows a curved trajectory, with motion in both the $x$ and $y$ directions, as it approaches its equilibrium position. Fig.~\ref{fig:ellipse_trajectory}(b) shows the corresponding particle orientation along the trajectory. The major axis gradually rotates toward an orientation perpendicular to the nodal plane as the particle migrates toward the pressure node.

Overall, these results demonstrate the capability of the proposed framework to capture the dynamics of asymmetric particles in acoustically actuated flows, including the coupled translational and rotational motion arising from fluid--particle interactions.



\section{Conclusions and Discussion}

In this study, we presented a volume penalization framework for simulating the dynamics of immersed rigid particles in acoustically actuated flows. The immersed solid is represented through a Brinkman penalization term incorporated directly into the compressible Navier--Stokes equations posed over a unified fluid--solid domain. A perturbation decomposition separates the nonlinear response into a first-order oscillatory component and a second-order time-averaged component, resulting in two sequentially coupled linear systems. Consistent expressions for the structural velocities, comprising rigid-body and Stokes-drift contributions, entering the penalization forcing are derived at both perturbation orders. To reduce numerical inaccuracies associated with evaluating hydrodynamic forces and torques across the diffuse fluid--solid interface, we employ a contour-integration approach in which the integration surface is displaced into the surrounding fluid. The particle position and orientation are then advanced by time-integrating the translational and rotational velocities obtained from the first- and second-order solutions.

We first applied the framework to a circular particle undergoing purely translational motion. The simulations showed that the acoustic streaming field around a freely moving particle differs substantially, both qualitatively and quantitatively, from that around a fixed particle. This result demonstrates that neglecting particle motion can lead to significant differences in the predicted streaming dynamics and, consequently, in the interpretation of the coupled fluid--solid response. We further showed that the influence of particle density on the resulting dynamics is more complex than that suggested by the acoustic contrast factor alone, providing additional motivation for explicitly resolving fluid--solid coupling rather than inferring particle motion through post-processing of particle-free solutions. The perturbation-based formulation also permits particle advancement using time steps orders of magnitude larger than those required to directly resolve the acoustic oscillations, enabling efficient simulations over the much longer time scales associated with particle migration. Simulations of a small particle further showed that its trajectory closely follows the underlying streaming pattern, demonstrating that the framework captures both radiation-dominated and streaming-dominated regimes. Finally, simulations of an elliptical particle demonstrated the ability of the framework to capture coupled translational and rotational dynamics within the same formulation.

The present study is restricted to two-dimensional simulations of a single rigid particle, whereas many acoustofluidic applications involve multiple particles for which hydrodynamic and acoustic interactions can strongly influence the collective dynamics. Extension to multiparticle systems could be achieved by introducing separate indicator functions for the individual particles within the volume penalization framework and represents a natural next step. Another important extension is the relaxation of the rigidity constraint to accommodate deformable bodies, enabling simulations of soft particles and biological cells whose shapes evolve in response to acoustic and hydrodynamic stresses. Acoustofluidic systems also frequently involve gas bubbles, whose nonlinear oscillations introduce additional multiphase and compressibility effects. Incorporating such dynamics through an appropriate multiphase compressible formulation would further broaden the applicability of the present approach. More broadly, the framework developed here provides a foundation for simulations of acoustically driven flows involving multiple immersed bodies with increasingly complex geometries, material responses, and interfacial dynamics.

\section*{Acknowledgements}
This work was supported, in part, by the National Science Foundation (OAC-1931368, CBET-2234387, CBET-2407937, CBET-2407938, CBET-2543194) and the American Heart Association (23CDA1048343).

\section*{Appendix}
\begin{appendix}

\section{Contour evaluation of first-order force and torque}
\label{app:fo_contour_integration}
\begin{figure}
    \centering
    \includegraphics[width=\linewidth]{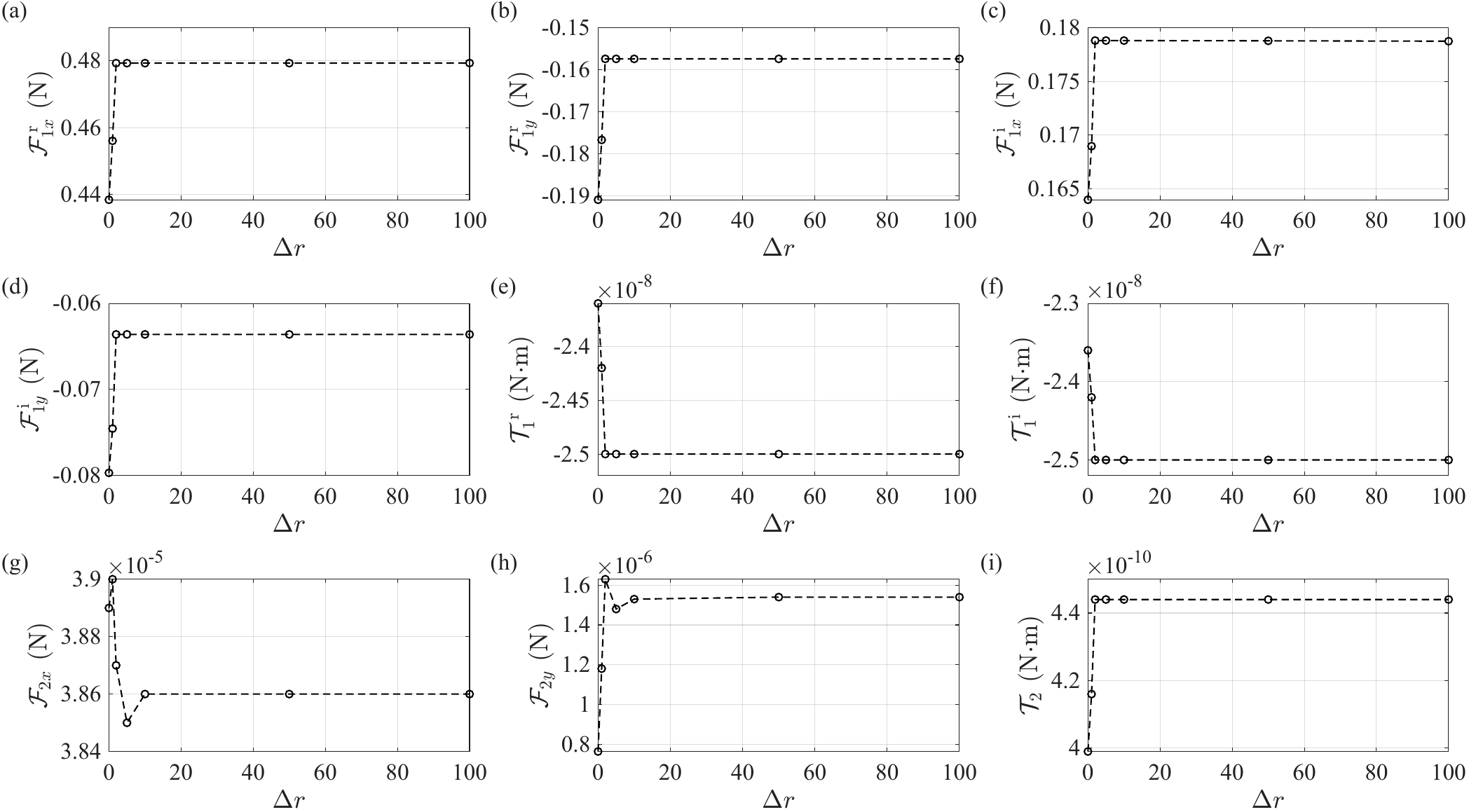}
    \caption{Convergence of first- and second-order forces and torques for an elliptical particle as considered in Sec.~\ref{sec:result:translation_rotation}, when the ellipse is in its initial position and orientation. First-order force and torque evaluated at different circular contours encompassing the solid. $\Delta r = (r-a)/h$, in which $a$ is semi-major axis of the ellipse. $\Delta r = 0$ denotes the integration circle touching fluid-solid interface (zero contour of diffused interface)}
    \label{fig:fo_contour_integration}
\end{figure}

The first- and second-order forces and torques are evaluated using the contour-integration formulation described in Sec.~\ref{sec:force_and_torque} for the case presented in Sec.~\ref{sec:result:translation_rotation}, with the ellipse in its initial position and orientation. The contour corresponding to $\Delta r = (r-a)/h = 0$ is referred to as the zero contour, where $a$ is the semi-major axis and $h$ is the grid spacing. This circular integration contour overlaps the fluid--solid interface at the ends of the ellipse's major axis. Fig.~\ref{fig:fo_contour_integration} illustrates the contour-integration results obtained using contours at different radial distances from the zero contour. The contour integration is repeated at distances of $1h$, $2h$, $5h$, $10h$, $50h$, and $100h$ outward from the zero contour. The results show that the first-order forces and torques converge for contours located beyond $2h$ from the zero contour, whereas the second-order forces and torques converge at a distance of $10h$ from the zero-contour. Accordingly, all force and torque evaluations throughout this work are performed using a contour located $100h$ radially outward from the zero contour.
 
%

\section{Rate of change of angular momentum of a rigid body defined over a moving volume} \label{sec_ang_mom}

Let $\volume_\mathrm{s}(t)$ denote the spatial volume occupied by the rigid body, with
boundary \(S(t)=\partial \volume_\mathrm{s}(t)\). The angular momentum about a fixed
inertial/laboratory frame origin $O$ is

\begin{equation}
\mathbf H_O(t)
=
\iiint_{\volume_\mathrm{s}(t)}
\rho\,\mathbf x\times \mathbf v\, \d V.
\end{equation}
For a moving control volume, the Leibniz rule (Reynolds transport theorem)
for a vector field \(\mathbf f(\mathbf x,t)\) is

\begin{equation}
\frac{\d}{ \d t}
\iiint_{\volume_\mathrm{s}(t)} \mathbf f\, \d V
=
\iiint_{\volume_\mathrm{s}(t)}
\frac{\partial \mathbf f}{\partial t}\, \d V
+
\iint_{S(t)}
\mathbf f\,(\mathbf w\cdot\mathbf n)\, \d S
\end{equation}
in which \(\mathbf w\) is the velocity of the moving boundary.
For a material rigid body, the boundary moves with the material velocity, so that $\mathbf w=\mathbf v$. Taking
$\mathbf f = \rho\,\mathbf x\times\mathbf v$, and assuming \(\rho\) is constant, we obtain

\begin{equation}
\frac{\d\mathbf H_O}{\d t}
=
\iiint_{\volume_\textrm{s}(t)}
\rho
\frac{\partial}{\partial t}
\left(
\mathbf x\times\mathbf v
\right)\, \d V
+
\iint_{S(t)}
\rho
\left(
\mathbf x\times\mathbf v
\right)
(\mathbf v\cdot\mathbf n)\,\d S.
\end{equation}
Since \(\mathbf x\) is the Eulerian spatial position,
\begin{equation}
\left.
\frac{\partial \mathbf x}{\partial t}
\right|_{\mathbf x}
=0,
\end{equation}
and therefore
\begin{equation}
\frac{\partial}{\partial t}
\left(
\mathbf x\times\mathbf v
\right)
=
\mathbf x\times
\frac{\partial\mathbf v}{\partial t}.
\end{equation}
Hence
\begin{equation}
\frac{\d\mathbf H_O}{\d t}
=
\iiint_{\volume_\mathrm{s}(t)}
\rho
\mathbf x\times
\frac{\partial\mathbf v}{\partial t}\, \d V
+
\iint_{S(t)}
\rho
\left(
\mathbf x\times\mathbf v
\right)
(\mathbf v\cdot\mathbf n)\, \d S.
\end{equation}
Using the tensor divergence theorem,
\begin{equation}
\iint_{S(t)}
\left(
\mathbf x\times\mathbf v
\right)
(\mathbf v\cdot\mathbf n)\, \d S
=
\iiint_{\volume_\mathrm{s}(t)}
\nabla\cdot
\left[
(\mathbf x\times\mathbf v)\otimes\mathbf v
\right]\, \d V.
\end{equation}
For any vector fields \(\mathbf a\) and \(\mathbf b\),
\begin{equation}
\nabla\cdot(\mathbf a\otimes\mathbf b)
=
(\mathbf b\cdot\nabla)\mathbf a
+
\mathbf a(\nabla\cdot\mathbf b).
\end{equation}
Thus,
\begin{equation}
\nabla\cdot
\left[
(\mathbf x\times\mathbf v)\otimes\mathbf v
\right]
=
(\mathbf v\cdot\nabla)
(\mathbf x\times\mathbf v)
+
(\mathbf x\times\mathbf v)
(\nabla\cdot\mathbf v).
\end{equation}
For rigid-body motion,
\begin{equation}
\mathbf v
=
\mathbf U_{\rm c}
+
\boldsymbol\Omega\times\mathbf r,
\qquad
\mathbf r=\mathbf x-\mathbf x_{\rm c},
\end{equation}
in which $ \mathbf U_{\rm c}\equiv \U$ denotes the translational velocity of the body, and the velocity field is divergence-free, $\nabla\cdot\mathbf v=0.$ Therefore,
\begin{equation}
\frac{\d \mathbf H_O}{\d t}
=
\iiint_{\volume_\mathrm{s}(t)}
\rho
\left[
\mathbf x\times
\frac{\partial\mathbf v}{\partial t}
+
(\mathbf v\cdot\nabla)
(\mathbf x\times\mathbf v)
\right]\, \d V.
\end{equation}
Now,
\begin{align}
(\mathbf v\cdot\nabla)
(\mathbf x\times\mathbf v)
&=
\left[
(\mathbf v\cdot\nabla)\mathbf x
\right]\times\mathbf v
+
\mathbf x\times
\left[
(\mathbf v\cdot\nabla)\mathbf v
\right]
\\
&=
\mathbf v\times\mathbf v
+
\mathbf x\times
\left[
(\mathbf v\cdot\nabla)\mathbf v
\right]
\\
&=
\mathbf x\times
\left[
(\mathbf v\cdot\nabla)\mathbf v
\right].
\end{align}
Hence,
\begin{align}
\frac{\d\mathbf H_O}{\d t}
&=
\iiint_{\volume_\mathrm{s}(t)}
\rho
\mathbf x\times
\left[
\frac{\partial\mathbf v}{\partial t}
+
(\mathbf v\cdot\nabla)\mathbf v
\right]dV
\\
&=
\iiint_{\volume_\mathrm{s}(t)}
\rho\,
\mathbf x\times \DDD{\v}{t}\,\d V.
\end{align}
Since $\DDD{\v}{t} = \mathbf a$ is the acceleration of the body, we obtain
\begin{equation}
\frac{\d\mathbf H_O}{\d t}
=
\iiint_{\volume_\mathrm{s}(t)}
\rho\,
\mathbf x\times\mathbf a\, \d V.
\end{equation}
For a rigid body we write the position of a material point as
$\mathbf x = \mathbf x_{\rm c} +\mathbf r$, in which $\x_{\rm c}$ is the center of mass and
\begin{equation}
\iiint_{\volume_\mathrm{s}(t)} \rho\,\mathbf r\,\d V = 0.
\end{equation}
The rigid-body acceleration is
\begin{equation}
\mathbf a
=
\mathbf a_{\rm c}
+
\dot{\boldsymbol\Omega}\times\mathbf r
+
\boldsymbol\Omega\times
(\boldsymbol\Omega\times\mathbf r),
\end{equation}
in which $ \mathbf a_{\rm c}=\d\mathbf U_{\rm c}/\d t$ denotes the translational acceleration of the center of mass and 
the relation 
$
\dot{\mathbf r} =
\boldsymbol\Omega\times\mathbf r
$
has been used. Here, $\V{\Omega}$ is expressed in the fixed laboratory frame. Substitution gives
\begin{align}
\iiint_{\volume_\mathrm{s}(t)}
\rho\,\mathbf x\times\mathbf a\, \d V
=
\iiint_{\volume_\mathrm{s}(t)}
\rho
(\mathbf x_{\rm c} +\mathbf r)
\times
\Big[
\mathbf a_{\rm c}
+
\dot{\boldsymbol\Omega}\times\mathbf r
+
\boldsymbol\Omega\times
(\boldsymbol\Omega\times\mathbf r)
\Big]\, \d V. \label{eqn_H_ac}
\end{align}
Using the definitions
\begin{equation}
\iiint_{\volume_\textrm{s}(t)}\rho\, \d V = m,
\qquad
\iiint_{\volume_\textrm{s}(t)}\rho\,\mathbf r\, \d V= 0,
\end{equation}
all terms linear in \(\mathbf r\) vanish from Eq.~\eqref{eqn_H_ac}, yielding

\begin{align}
\iiint_{\volume_\textrm{s}(t)}
\rho\,\mathbf x\times\mathbf a\, \d V
={}&
m\,\mathbf x_{\rm c} \times\mathbf a_{\rm c} 
+
\iiint_{\volume_\textrm{s}(t)}
\rho
\mathbf r\times
(\dot{\boldsymbol\Omega}\times\mathbf r)\, \d V 
+
\iiint_{\volume_\textrm{s}(t)}
\rho
\mathbf r\times
\left[
\boldsymbol\Omega\times
(\boldsymbol\Omega\times\mathbf r)
\right]\, \d V. \label{eqn_int_accn}
\end{align}
Define the time-dependent inertia tensor in the fixed laboratory frame
\begin{equation}
\V{\mathcal{I}}_{\rm c}^{\rm lab}(t)
=
\iiint_{\volume_\textrm{s}(t)}
\rho
\left[
(\mathbf r\cdot\mathbf r)\mathbf I
-
\mathbf r\otimes\mathbf r
\right]\, \d V.
\end{equation}
%
Using the vector identity
\begin{equation}
\mathbf r\times(\mathbf q\times\mathbf r)
=
\left[
(\mathbf r\cdot\mathbf r)\mathbf I
-
\mathbf r\otimes\mathbf r
\right]\mathbf q,
\end{equation}
we find
\begin{equation}
\iiint_{\volume_\textrm{s}(t)}
\rho
\mathbf r\times
(\dot{\boldsymbol\Omega}\times\mathbf r)\, \d V
= \V{\mathcal{I}}_{\rm c}^{\rm lab} \dot{\boldsymbol\Omega}. \label{eqn_I_dotomega}
\end{equation}
Similarly, the last term on the right-hand side of Eq.~\eqref{eqn_int_accn} can be written as 
\begin{equation}
\iiint_{\volume_\textrm{s}(t)} \rho \mathbf r\times \left[ \boldsymbol\Omega\times (\boldsymbol\Omega\times\mathbf r) \right] \d V = \boldsymbol\Omega\times ( \V{\mathcal{I}}_{\rm c}^{\rm lab}\boldsymbol\Omega). \label{eqn_omega_I_omega}
\end{equation}
Note that in two dimensions, $\boldsymbol\Omega\times
(\V{\mathcal{I}}_{\rm c}\boldsymbol\Omega) \equiv 0$.
In light of Eqs.~\eqref{eqn_I_dotomega} and~\eqref{eqn_omega_I_omega}, Eq.~\eqref{eqn_int_accn} reads as
\begin{equation}
\frac{\d\mathbf H_O}{\d t} 
=
m\,\mathbf x_{\rm c}\times\mathbf a_{\rm c}
+
\V{\mathcal{I}}_{\rm c}^{\rm lab} \, \dot{\boldsymbol\Omega}
+
\boldsymbol\Omega\times
(\V{\mathcal{I}}_{\rm c}^{\rm lab} \,\boldsymbol\Omega). \label{eqn_dHdt_O}
\end{equation}
in which all vector components are expressed in the fixed laboratory frame.  
Since the inertia tensor expressed in the laboratory frame varies with
the orientation of the body, it is convenient to relate it to the
time-independent inertia tensor expressed in the body-fixed frame.
The two representations are related through the time-dependent rotation
matrix $\V{\mathcal R}(t)$ as
\begin{equation}
    \bm{\mathcal I}_{\rm c}^{\rm lab}(t)
    =
    \V{\mathcal R}(t)
    \bm{\mathcal I}_{\rm c}^{\rm body}
    \V{\mathcal R}^{\intercal}(t).
    \label{eqn:Ilab_Ibody}
\end{equation}

\section{Rate of change of linear momentum of a rigid body defined over a moving volume} \label{sec_lin_mom}
Starting from the rigid-body velocity field, $\mathbf{v} =
\mathbf{U}_{\rm c} + \boldsymbol{\Omega}\times\mathbf{r}$,
the total linear momentum of the rigid body is
\begin{align}
\bm{\mathcal{P}}
&=
\iiint_{\volume_\textrm{s}(t)}
\rho\,\mathbf{v}\, \d V \nonumber \\
&=
\iiint_{\volume_\textrm{s}(t)}
\rho
\left(
\mathbf{U}_{\rm c}
+
\boldsymbol{\Omega}\times\mathbf{r}
\right) \d V \nonumber \\
&=
\mathbf{U}_{\rm c}
\iiint_{\volume_\textrm{s}(t)} \rho\, \d V
+
\boldsymbol{\Omega}\times
\iiint_{\volume_\textrm{s}(t)} \rho\,\mathbf{r}\, \d V.
\end{align}
By definition of the center of mass $\iiint_{\volume_\textrm{s}(t)}
\rho\,\mathbf{r}\, \d V
= 0,
$
and the total mass 
$
m
=
\iiint_{\volume_\textrm{s}(t)}\rho\,\d V$,  the linear momentum $\bm{\mathcal{P}}$ is simply
\begin{equation}
\bm{\mathcal{P}}
=
m\,\mathbf{U}_{\rm c}.
\end{equation}
Taking the time derivative, and noting that the mass of the rigid body
is constant, i.e., 
$
\frac{\d}{\d t}\iiint_{\volume_\textrm{s}(t)} \rho\, \d V = 0
$,
gives
\begin{align}
\frac{\d\bm{\mathcal{P}}}{\d t}
&=
\frac{\d}{\d t}
\left(
m\mathbf{U}_{\rm c}
\right) =
m\frac{\d \mathbf{U}_{\rm c}}{\d t}
= m \, \mathbf{a}_{\rm c}.
\end{align}

\section{Second-order mean angular velocity of the rigid body}
\label{sec:various_omegas_orientation}

\subsection{Derivation using rotation tensors}

The instantaneous orientation of the immersed rigid body is
described by a proper orthogonal rotation tensor
$\mathbf{Q}(t)$, satisfying
\begin{equation}
    \mathbf{Q}^{\intercal}\mathbf{Q}=\mathbf{I}.
\end{equation}
The exact rotational kinematics are
\begin{equation}
    \dot{\mathbf{Q}}
    =
    [\boldsymbol{\mathrm{\Omega}}]_{\times}\mathbf{Q},
    \label{eq:exact_rotation_kinematics}
\end{equation}
in which $[\boldsymbol{\mathrm{\Omega}}]_{\times}$ denotes
the skew-symmetric cross-product matrix associated with
the instantaneous angular velocity
$\boldsymbol{\mathrm{\Omega}}$.

To distinguish the rapidly oscillating acoustic rotation
from the slowly evolving mean orientation, we introduce
\begin{equation}
    \mathbf{Q}(t)
    =
    \mathbf{P}(t)\overline{\mathbf{Q}}(t),
    \label{eq:rotation_decomposition}
\end{equation}
in which $\mathbf{P}$ describes the small acoustic periodic rotation,
while $\overline{\mathbf{Q}}$ describes the evolving mean
orientation.
The decomposition is chosen such that the
rotation-vector perturbations associated with
$\mathbf{P}$ are bounded and periodic, with zero mean
through second order. The angular velocity governing the mean orientation
is defined through
\begin{equation}
    \dot{\overline{\mathbf{Q}}}
    =
    [\boldsymbol{\mathrm{\Omega}}_{\mathrm{mean}}]_{\times}
    \overline{\mathbf{Q}}.
    \label{eq:mean_rotation_kinematics}
\end{equation}
In what follows, we derive an expression for $\boldsymbol{\mathrm{\Omega}}_{\mathrm{mean}}$ in terms of first- and second-order rotational velocities of the body. 
Differentiating Eq.~\eqref{eq:rotation_decomposition} yields
\begin{equation}
    \dot{\mathbf{Q}}
    =
    \dot{\mathbf{P}}\overline{\mathbf{Q}}
    +
    \mathbf{P}\dot{\overline{\mathbf{Q}}}.
\end{equation}
Multiplication by $\mathbf{Q}^{\intercal}$ gives
\begin{equation}
    \dot{\mathbf{Q}}\mathbf{Q}^{\intercal}
    =
    \dot{\mathbf{P}}\mathbf{P}^{\intercal}
    +
    \mathbf{P}
    \dot{\overline{\mathbf{Q}}} \;
    \overline{\mathbf{Q}}^{\,\intercal}
    \mathbf{P}^{\intercal}.
\end{equation}
Using the definitions of the instantaneous and mean
angular velocities,
\begin{equation}
    [\boldsymbol{\mathrm{\Omega}}]_{\times}
    =
    \dot{\mathbf{P}}\mathbf{P}^{\intercal}
    +
    \mathbf{P}
    [\boldsymbol{\mathrm{\Omega}}_{\mathrm{mean}}]_{\times}
    \mathbf{P}^{\intercal}.
    \label{eq:angular_velocity_decomposition}
\end{equation}
As done in Sec.~\ref{sec_kinematics_osc_body}, the instantaneous angular velocity is expanded as
\begin{equation}
    \boldsymbol{\mathrm{\Omega}}
    =
    \boldsymbol{\mathrm{\Omega}}_1
    +
    \boldsymbol{\mathrm{\Omega}}_2
    +
    O(\epsilon^3),
\end{equation}
in which $\boldsymbol{\mathrm{\Omega}}_1=O(\epsilon)$
and $\boldsymbol{\mathrm{\Omega}}_2=O(\epsilon^2)$.
The mean angular velocity is also second order,
\begin{equation}
    \boldsymbol{\mathrm{\Omega}}_{\mathrm{mean}}
    =
    O(\epsilon^2).
\end{equation}
Since $\mathbf{P}=\mathbf{I}+O(\epsilon)$,
\begin{equation}
    \mathbf{P}
    [\boldsymbol{\mathrm{\Omega}}_{\mathrm{mean}}]_{\times}
    \mathbf{P}^{\intercal}
    =
    [\boldsymbol{\mathrm{\Omega}}_{\mathrm{mean}}]_{\times}
    +
    O(\epsilon^3).
\end{equation}
Consequently,
\begin{equation}
    [\boldsymbol{\mathrm{\Omega}}]_{\times}
    =
    \dot{\mathbf{P}}\mathbf{P}^{\intercal}
    +
    [\boldsymbol{\mathrm{\Omega}}_{\mathrm{mean}}]_{\times}
    +
    O(\epsilon^3).
    \label{eq:angular_velocity_second_order}
\end{equation}
The acoustic rotation can also be represented using the exponential map
\begin{equation}
    \mathbf{P}
    =
    \exp([\boldsymbol{\mathrm{\theta}}]_{\times}),
\end{equation}
which provides an exact representation of a three-dimensional rigid-body rotation and is equivalent to Rodrigues' rotation formula.  Expanding $\boldsymbol{\mathrm{\theta}}$ in powers of $\epsilon$ yields 
\begin{equation}
    \boldsymbol{\mathrm{\theta}}
    =
    \boldsymbol{\mathrm{\theta}}_1
    +
    \boldsymbol{\mathrm{\theta}}_2
    +
    O(\epsilon^3),
\end{equation}
in which $\boldsymbol{\mathrm{\theta}}_1=O(\epsilon)$
and $\boldsymbol{\mathrm{\theta}}_2=O(\epsilon^2)$
are bounded, periodic rotation-vector perturbations.
Introducing the skew-symmetric tensor
\begin{equation}
    \mathbf{A}
    =
    [\boldsymbol{\mathrm{\theta}}]_{\times},
\end{equation}
the matrix exponential is expanded as
\begin{equation}
    \mathbf{P}
    =
    \mathbf{I}
    +
    \mathbf{A}
    +
    \frac{1}{2}\mathbf{A}^{2}
    +
    O(\epsilon^3).
\end{equation}
Since $\mathbf{A}^{\intercal}=-\mathbf{A}$,
\begin{equation}
    \mathbf{P}^{\intercal}
    =
    \mathbf{I}
    -
    \mathbf{A}
    +
    \frac{1}{2}\mathbf{A}^{2}
    +
    O(\epsilon^3).
\end{equation}
Differentiating the expansion of $\mathbf{P}$ gives
\begin{equation}
    \dot{\mathbf{P}}
    =
    \dot{\mathbf{A}}
    +
    \frac{1}{2}\dot{\mathbf{A}}\mathbf{A}
    +
    \frac{1}{2}\mathbf{A}\dot{\mathbf{A}}
    +
    O(\epsilon^3).
\end{equation}
Multiplying by $\mathbf{P}^{\intercal}$ and retaining terms
through second order gives
\begin{equation}
    \dot{\mathbf{P}}\mathbf{P}^{\intercal}
    =
    \dot{\mathbf{A}}
    +
    \frac{1}{2}\mathbf{A}\dot{\mathbf{A}}
    -
    \frac{1}{2}\dot{\mathbf{A}}\mathbf{A}
    +
    O(\epsilon^3).
    \label{eq:fast_rotation_matrix_product}
\end{equation}
For an arbitrary vector $\mathbf{b}$,
\begin{equation}
    \mathbf{A}\dot{\mathbf{A}}\mathbf{b}
    =
    \boldsymbol{\mathrm{\theta}}
    \times
    (\dot{\boldsymbol{\mathrm{\theta}}}
    \times\mathbf{b}),
\end{equation}
whereas
\begin{equation}
    \dot{\mathbf{A}}\mathbf{A}\mathbf{b}
    =
    \dot{\boldsymbol{\mathrm{\theta}}}
    \times
    (\boldsymbol{\mathrm{\theta}}\times\mathbf{b}).
\end{equation}
Using the vector triple-product identity,
\begin{equation}
    (\mathbf{A}\dot{\mathbf{A}}
    -
    \dot{\mathbf{A}}\mathbf{A})\mathbf{b}
    =
    (\boldsymbol{\mathrm{\theta}}
    \times\dot{\boldsymbol{\mathrm{\theta}}})
    \times\mathbf{b}.
\end{equation}
Since this relation holds for any vector $\mathbf{b}$,
\begin{equation}
    \mathbf{A}\dot{\mathbf{A}}
    -
    \dot{\mathbf{A}}\mathbf{A}
    =
    [\boldsymbol{\mathrm{\theta}}
    \times\dot{\boldsymbol{\mathrm{\theta}}}]_{\times}.
\end{equation}
Substitution into
Eq.~\eqref{eq:fast_rotation_matrix_product} gives
\begin{equation}
    \dot{\mathbf{P}}\mathbf{P}^{\intercal}
    =
    \left[
    \dot{\boldsymbol{\mathrm{\theta}}}
    +
    \frac{1}{2}
    \boldsymbol{\mathrm{\theta}}
    \times\dot{\boldsymbol{\mathrm{\theta}}}
    \right]_{\times}
    +
    O(\epsilon^3).
    \label{eq:fast_rotation_angular_velocity}
\end{equation}
Substituting the perturbation expansion of
$\boldsymbol{\mathrm{\theta}}$ and retaining terms
through second order yields
\begin{equation}
    \dot{\mathbf{P}}\mathbf{P}^{\intercal}
    =
    \left[
    \dot{\boldsymbol{\mathrm{\theta}}}_1
    +
    \dot{\boldsymbol{\mathrm{\theta}}}_2
    +
    \frac{1}{2}
    \boldsymbol{\mathrm{\theta}}_1
    \times\dot{\boldsymbol{\mathrm{\theta}}}_1
    \right]_{\times}
    +
    O(\epsilon^3).
\end{equation}
Combining with
Eq.~\eqref{eq:angular_velocity_second_order}
and matching first-order terms yields
\begin{equation}
    \dot{\boldsymbol{\mathrm{\theta}}}_1
    =
    \boldsymbol{\mathrm{\Omega}}_1.
    \label{eq:theta1_Omega1}
\end{equation}
At second order,
\begin{equation}
    \boldsymbol{\mathrm{\Omega}}_2
    =
    \dot{\boldsymbol{\mathrm{\theta}}}_2
    +
    \frac{1}{2}
    \boldsymbol{\mathrm{\theta}}_1
    \times\boldsymbol{\mathrm{\Omega}}_1
    +
    \boldsymbol{\mathrm{\Omega}}_{\mathrm{mean}}.
    \label{eq:Omega2_mean_relation}
\end{equation}
We next average Eq.~\eqref{eq:Omega2_mean_relation}
over an acoustic period. Since the second-order fast
rotation is also assumed periodic (to ensure that $\V{P}$ is periodic),
\begin{equation}
    \langle
    \dot{\boldsymbol{\mathrm{\theta}}}_2
    \rangle
    = 0.
\end{equation}
Consequently,
\begin{equation}
    \left\langle
    \boldsymbol{\mathrm{\Omega}}_2
    \right\rangle
    =
    \frac{1}{2}
    \left\langle
    \boldsymbol{\mathrm{\theta}}_1
    \times\boldsymbol{\mathrm{\Omega}}_1
    \right\rangle
    +
    \boldsymbol{\mathrm{\Omega}}_{\mathrm{mean}}.
\end{equation}
Rearranging the above equation gives the mean angular velocity $\boldsymbol{\mathrm{\Omega}}_{\mathrm{mean}}$ needed to update the orientation of the body
\begin{equation}
    \boldsymbol{\mathrm{\Omega}}_{\mathrm{mean}}
    =
    \left\langle
    \boldsymbol{\mathrm{\Omega}}_2
    \right\rangle
    -
    \frac{1}{2}
    \left\langle
    \boldsymbol{\mathrm{\theta}}_1
    \times\boldsymbol{\mathrm{\Omega}}_1
    \right\rangle.
    \label{eq:mean_angular_velocity}
\end{equation}
The last term of Eq.~\eqref{eq:mean_angular_velocity} can be obtained as follows. The first-order angular velocity is represented using
the complex harmonic convention
\begin{equation}
    \boldsymbol{\mathrm{\Omega}}_1(t)
    =
    \operatorname{Re}
    \left\{
    \widehat{\boldsymbol{\mathrm{\Omega}}}_1
    e^{\iota\omega t}
    \right\},
\end{equation}
in which $\omega$ is the acoustic
angular frequency. The complex amplitude is
\begin{equation}
    \widehat{\boldsymbol{\mathrm{\Omega}}}_1
    =
    \boldsymbol{\mathrm{\Omega}}_1^{\rm r}
    +
    \iota \boldsymbol{\mathrm{\Omega}}_1^{\rm i}.
\end{equation}
Using Eq.~\eqref{eq:theta1_Omega1},
\begin{equation}
    \widehat{\boldsymbol{\mathrm{\theta}}}_1
    =
    \frac{
    \widehat{\boldsymbol{\mathrm{\Omega}}}_1
    }{\iota \omega}.
\end{equation}
Therefore,
\begin{equation}
    \boldsymbol{\mathrm{\theta}}_1^{\rm r}
    =
    \frac{
    \boldsymbol{\mathrm{\Omega}}_1^{\rm i}
    }{\omega},
\end{equation}
and
\begin{equation}
    \boldsymbol{\mathrm{\theta}}_1^{\rm i}
    =
    -\frac{
    \boldsymbol{\mathrm{\Omega}}_1^{\rm r}
    }{\omega}.
\end{equation}
The acoustic-period average is
\begin{equation}
    \left\langle
    \boldsymbol{\mathrm{\theta}}_1
    \times\boldsymbol{\mathrm{\Omega}}_1
    \right\rangle
    =
    \frac{1}{2}
    \left(
    \boldsymbol{\mathrm{\theta}}_1^{\rm r}
    \times\boldsymbol{\mathrm{\Omega}}_1^{\rm r}
    +
    \boldsymbol{\mathrm{\theta}}_1^{\rm i}
    \times\boldsymbol{\mathrm{\Omega}}_1^{\rm i}
    \right).
\end{equation}
Substitution gives
\begin{equation}
    \left\langle
    \boldsymbol{\mathrm{\theta}}_1
    \times\boldsymbol{\mathrm{\Omega}}_1
    \right\rangle
    =
    -\frac{1}{\omega}
    \boldsymbol{\mathrm{\Omega}}_1^{\rm r}
    \times\boldsymbol{\mathrm{\Omega}}_1^{\rm i}.
\end{equation}
Thus, the effective mean angular velocity becomes
\begin{equation}
    \boldsymbol{\mathrm{\Omega}}_{\mathrm{mean}}
    =
    \boldsymbol{\mathrm{\Omega}}_2
    +
    \frac{1}{2\omega}
    \boldsymbol{\mathrm{\Omega}}_1^{\rm r}
    \times\boldsymbol{\mathrm{\Omega}}_1^{\rm i},
    \label{eq:mean_angular_velocity_harmonic}
\end{equation}
in which $\langle \cdot \rangle$ has been dropped from the second-order rotational velocity $\V{\Omega}_2$, which is already assumed to be a time-averaged quantity.

\subsection{Derivation using the Lagrangian velocity of the body}
It is also instructive to see the derivation of the mean angular velocity of the body (Eq.~\eqref{eq:mean_angular_velocity}) directly from its second-order Lagrangian velocity $\v_2^{\rm L}$. 
Taking the time-average of Eq.~\eqref{eq:v2b_lagrangian_theta} yields
\begin{equation}
    \left\langle\mathbf{v}_{2}^{\rm L}\right\rangle
    =
    \mathbf{U}_2
    +
    \bm{\mathrm{\Omega}}_2
    \times\overline{\mathbf{r}}
    +
    \left\langle
    \bm{\mathrm{\Omega}}_1
    \times
    \left(
    \bm{\mathrm{\theta}}_1
    \times\overline{\mathbf{r}}
    \right)
    \right\rangle.
    \label{eq:v2b_lagrangian_average}
\end{equation}
To identify the rotational contribution associated
with the nonlinear term in
Eq.~\eqref{eq:v2b_lagrangian_average},
we recall the cross-product matrix
$[\mathbf{a}]_{\times}$, defined by
\begin{equation}
    [\mathbf{a}]_{\times}\mathbf{b}
    =
    \mathbf{a}\times\mathbf{b}.
\end{equation}
The nonlinear rotational contribution can then be
written as
\begin{equation}
    \left\langle
    \bm{\mathrm{\Omega}}_1
    \times
    \left(
    \bm{\mathrm{\theta}}_1
    \times\overline{\mathbf{r}}
    \right)
    \right\rangle
    =
    \mathbf{M}\overline{\mathbf{r}},
\end{equation}
in which
\begin{equation}
    \mathbf{M}
    =
    \left\langle
    [\bm{\mathrm{\Omega}}_1]_{\times}
    [\bm{\mathrm{\theta}}_1]_{\times}
    \right\rangle.
    \label{eq:M_definition}
\end{equation}
The tensor $\mathbf{M}$ can be decomposed into
its symmetric and antisymmetric parts:
\begin{equation}
    \mathbf{M}
    =
    \mathbf{M}_{\mathrm{sym}}
    +
    \mathbf{M}_{\mathrm{skew}},
\end{equation}
in which
\begin{subequations}
\begin{alignat}{2}
    \mathbf{M}_{\mathrm{sym}}
    &=
    \frac{1}{2}
    \left(
    \mathbf{M}+\mathbf{M}^{\intercal}
    \right), \label{eq:M_symmetric} \\
    \mathbf{M}_{\mathrm{skew}}
    &=
    \frac{1}{2}
    \left(
    \mathbf{M}-\mathbf{M}^{\intercal}
    \right).
    \label{eq:M_skew}
\end{alignat}
\end{subequations}
We first examine the symmetric part. Since
cross-product matrices are skew symmetric,
\begin{equation}
    [\bm{\mathrm{\theta}}_1]_{\times}^{\intercal}
    =
    -[\bm{\mathrm{\theta}}_1]_{\times},
\end{equation}
and similarly for
$[\bm{\mathrm{\Omega}}_1]_{\times}$.
Therefore,
\begin{equation}
    \mathbf{M}^{\intercal}
    =
    \left\langle
    [\bm{\mathrm{\theta}}_1]_{\times}
    [\bm{\mathrm{\Omega}}_1]_{\times}
    \right\rangle.
\end{equation}
Substitution into Eq.~\eqref{eq:M_symmetric}
gives
\begin{equation}
    \mathbf{M}_{\mathrm{sym}}
    =
    \frac{1}{2}
    \left\langle
    [\bm{\mathrm{\Omega}}_1]_{\times}
    [\bm{\mathrm{\theta}}_1]_{\times}
    +
    [\bm{\mathrm{\theta}}_1]_{\times}
    [\bm{\mathrm{\Omega}}_1]_{\times}
    \right\rangle.
    \label{eq:M_sym_expanded}
\end{equation}
Define
\begin{equation}
    \mathbf{B}_1
    =
    [\bm{\mathrm{\theta}}_1]_{\times}.
\end{equation}
Using the first-order rotational kinematics,
Eq.~\eqref{eq:theta1_Omega1},
\begin{equation}
    \dot{\mathbf{B}}_1
    =
    [\bm{\mathrm{\Omega}}_1]_{\times}.
\end{equation}
Eq.~\eqref{eq:M_sym_expanded} becomes
\begin{equation}
    \mathbf{M}_{\mathrm{sym}}
    =
    \frac{1}{2}
    \left\langle
    \dot{\mathbf{B}}_1\mathbf{B}_1
    +
    \mathbf{B}_1\dot{\mathbf{B}}_1
    \right\rangle 
    =     \frac{1}{2}
    \left\langle
    \frac{\d}{\d t}
    \left(
    \mathbf{B}_1^2
    \right)
    \right\rangle = \bm{0}.
\end{equation}
Therefore, $\mathbf{M}_{\mathrm{sym}} = \V{0}$ because $\mathbf{B}_1$ represents a periodic rotation. In other words, the symmetric part of the nonlinear
rotational contribution has zero acoustic-period
average.
We next consider the antisymmetric part.
Substituting Eq.~\eqref{eq:M_definition}
into Eq.~\eqref{eq:M_skew} gives
\begin{equation}
    \mathbf{M}_{\mathrm{skew}}
    =
    \frac{1}{2}
    \left\langle
    [\bm{\mathrm{\Omega}}_1]_{\times}
    [\bm{\mathrm{\theta}}_1]_{\times}
    -
    [\bm{\mathrm{\theta}}_1]_{\times}
    [\bm{\mathrm{\Omega}}_1]_{\times}
    \right\rangle.
\end{equation}
For arbitrary vectors $\mathbf{a}$ and
$\mathbf{b}$, ordinary matrix multiplication
gives
\begin{equation}
    [\mathbf{a}]_{\times}
    [\mathbf{b}]_{\times}
    =
    \mathbf{b}\mathbf{a}^{\intercal}
    -
    (\mathbf{a}\cdot\mathbf{b})\mathbf{I}.
\end{equation}
Reversing the multiplication order,
\begin{equation}
    [\mathbf{b}]_{\times}
    [\mathbf{a}]_{\times}
    =
    \mathbf{a}\mathbf{b}^{\intercal}
    -
    (\mathbf{a}\cdot\mathbf{b})\mathbf{I}.
\end{equation}
Subtracting the two expressions gives
\begin{equation}
    [\mathbf{a}]_{\times}
    [\mathbf{b}]_{\times}
    -
    [\mathbf{b}]_{\times}
    [\mathbf{a}]_{\times}
    =
    \mathbf{b}\mathbf{a}^{\intercal}
    -
    \mathbf{a}\mathbf{b}^{\intercal}.
\end{equation}
The right-hand side is the cross-product
matrix associated with $\mathbf{a}\times\mathbf{b}$:
\begin{equation}
    \mathbf{b}\mathbf{a}^{\intercal}
    -
    \mathbf{a}\mathbf{b}^{\intercal}
    =
    [\mathbf{a}\times\mathbf{b}]_{\times}.
\end{equation}
Therefore,
\begin{equation}
    \mathbf{M}_{\mathrm{skew}}
    =
    \frac{1}{2}
    \left[
    \left\langle
    \bm{\mathrm{\Omega}}_1
    \times\bm{\mathrm{\theta}}_1
    \right\rangle
    \right]_{\times} 
    = 
    -\frac{1}{2}
    \left[
    \left\langle
    \bm{\mathrm{\theta}}_1
    \times\bm{\mathrm{\Omega}}_1
    \right\rangle
    \right]_{\times}.
\end{equation}
Since $\mathbf{M}_{\mathrm{sym}}=\mathbf{0}$,
the nonlinear rotational contribution to the
averaged Lagrangian velocity is purely
antisymmetric and can therefore be expressed
as a rigid-body rotational velocity. Hence,
\begin{equation}
    \left\langle
    \mathbf{v}_{2}^{\rm L}
    \right\rangle
    =
    \mathbf{U}_2
    +
    \bm{\mathrm{\Omega}}_{\mathrm{mean}}
    \times\overline{\mathbf{r}},
    \label{eq:v2b_mean_effective}
\end{equation}
in which
\begin{equation}
    \bm{\mathrm{\Omega}}_{\mathrm{mean}}
    =
    \bm{\mathrm{\Omega}}_2
    -
    \frac{1}{2}
    \left\langle
    \bm{\mathrm{\theta}}_1
    \times\bm{\mathrm{\Omega}}_1
    \right\rangle.
    \label{eq:Omega_eff_lagrangian}
\end{equation}
Eq.~\eqref{eq:Omega_eff_lagrangian} is identical to Eq.~\eqref{eq:mean_angular_velocity}, but is obtained independently from the time-averaged second-order Lagrangian velocity of the rigid body. Furthermore, Eq.~\eqref{eq:v2b_mean_effective} recovers the familiar rigid-body velocity form, consisting of a translational contribution and a rotational contribution governed by the effective mean angular velocity. 

\end{appendix}


\bibliography{acoustic_bibliography.bib}
\end{document}